\documentclass[preprint,eqsecnum,prb]{revtex4}
\usepackage{epsfig,amsmath,bbm,graphicx}

\newcommand{\ee}{\end{equation}}
\newcommand{\be}{\begin{equation}}
\newcommand{\bea}{\begin{eqnarray}}
\newcommand{\eea}{\end{eqnarray}}
\newcommand{\bml}{\begin{subequations}} 
\newcommand{\eml}{\end{subequations}}

\begin{document}

\title{On the Accuracy of the Noninteracting Electron Approximation for 
Vibrationally Coupled Electron Transport}

\author{Haobin Wang}
\affiliation{Department of Chemistry, University of Colorado Denver,
Denver, CO 80217-3364, USA}

\author{Michael Thoss}
\affiliation{Institut  f\"ur Theoretische Physik und 
Interdisziplin\"ares Zentrum f\"ur Molekulare Materialien,
  Friedrich-Alexander-Universit\"at Erlangen-N\"urnberg,
  Staudtstr.\ 7/B2, D-91058, Germany}


\begin{abstract}
\baselineskip6mm
The accuracy of the noninteracting electron approximation is examined for a model of 
vibrationally coupled electron transport in single molecule junction.
In the absence of electronic-vibrational coupling, steady state transport in this model is described exactly by Landauer
theory.  Including coupling, both electronic-vibrational and 
vibrationally induced electron-electron correlation effects may contribute to the real time quantum
dynamics.  Using the multilayer multiconfiguration time-dependent Hartree (ML-MCTDH) theory
to describe nuclear dynamics exactly while maintaining the noninteracting electron 
approximation for the electronic dynamics, the correlation effects are analyzed in different
physical regimes. It is shown that although the noninteracting electron approximation 
may be reasonable for describing short time dynamics, it does not give the correct long
time limit for certain initial conditions.

\end{abstract}
\maketitle

\section{Introduction}

There is considerable interest in modeling charge transport in single-molecule 
junctions.\cite{ree97:252,joa00:541,Nitzan01,nit03:1384,Cuniberti05,Selzer06,Venkataraman06,Chen07,Galperin08b,Cuevas10}  From a practical perspective, this may provide insight for the development
of molecular electronic devices.  A variety of experimental techniques, such 
electromigration, mechanically controllable break junctions, and scanning tunneling microscopy 
have been employed to study molecular junctions.\cite{ree97:252,par00:57,cui01:571,par02:722,smi02:906,rei02:176804,xu03:1221,qiu04:206102,liu04:11371,elb05:8815,Ogawa07,Schulze08,Pump08,Leon08,Osorio10,Martin10}  
In contrast to macroscopic conductors, molecular junctions typically have nonlinear current-voltage characteristics, which often show fine structures that reveal 
molecular details such as positions of molecular orbitals and vibrational signatures. From
a more fundamental point of view, the experiments have also revealed many interesting transport 
phenomena such as Coulomb blockade,\cite{par02:722} Kondo effect,\cite{lia02:725} negative 
differential resistance,\cite{che99:1550,Gaudioso00,Osorio10} switching and 
hysteresis,\cite{blu05:167,Riel06,Choi06} and quantum interference.\cite{Ballmann12,Guedon_Molen_12,Vazquez_Venkatamaran12} These findings have stimulated the development of
physical theories and simulation techniques that can be used to rationalize experimental results
and make predictions for improved designs of molecular junctions.

A useful approach for a qualitative modeling of the conductance in molecular junctions is Landauer theory.\cite{Landauer57,Cuevas10}
For noninteracting systems, such as, e.g., tight-binding based models of molecular junctions, it provides an exact description of steady state transport. However, it does not include correlation effects due to electron-electron or electronic-vibrational coupling.
To describe electron transport with electronic-vibrational interaction, more elaborate approximate theories have been used, such as the scattering
theory,\cite{Bonca95,Ness01,Cizek04,Cizek05,Toroker07,Benesch08,Zimbovskaya09,Seidemann10} 
nonequilibrium Green's function (NEGF)
approaches,\cite{Flensberg03,Mitra04,Galperin06,Ryndyk06,Frederiksen07,Tahir08,Haertle08,Stafford09,Haertle09} 
and master equation methods.\cite{May02,Mitra04,Lehmann04,Pedersen05,Harbola06,Zazunov06,Siddiqui07,Timm08,May08,May08b,Leijnse09,Esposito09,Haertle11} Furthermore,  numerically exact simulation methods have been developed such
as path integral,\cite{muh08:176403,wei08:195316,Segal10} real-time quantum Monte 
Carlo,\cite{Werner09,Schiro09} and numerical renormalization group approaches,\cite{and08:066804} 
the multilayer multiconfiguration time-dependent Hartree theory in second quantization 
representation (ML-MCTDH-SQR),\cite{wan09:024114,wan11:244506,wan13:134704,wan13:7431} as well as combinations of the latter method with reduced density matrix theory.\cite{wil13:045137} 
In  contrast to mesoscopic systems, molecular junctions often exhibit strong electronic-vibrational coupling and, therefore, the vibrations have to be included in the theoretical treatment. This coupling may give rise to substantial current-induced vibrational excitation and 
thus may cause heating and possible breakage of the molecular junction.
 The all-importance of vibrational effects in molecular junctions have also been confirmed by a variety of experiments.\cite{sti98:1732,par00:57,smi02:906,qiu04:206102,Natelson04,kus04:639,pas05:203,Sapmaz06,Thijssen06,Parks07,Boehler07,Leon08,Huang06,Natelson08,Ioffe08,Huettel09,Ballmann10,Osorio10,Secker11,Ballmann13}

Despite the importance of including the vibrations in electron transport through molecular 
junctions, a theoretical description that is both accurate and efficient still remains 
a challenging task. Numerically exact simulation methods are limited to certain physical
regimes and a small size of the molecular system.  Approximate theories can handle
larger systems but nevertheless involve significant approximations. For example,  NEGF methods 
and master equation approaches are usually based on (self-consistent) perturbation theory 
and/or employ factorization schemes. Scattering theory approaches to vibrationally coupled 
electron transport, on the other hand, neglect vibrational nonequilibrium effects and are 
limited to the treatment of a small number of vibrational degrees of freedom. It is thus
desirable to combine the above two strategies in practical applications. One may use numerically
exact methods to gauge the accuracy of approximate theories in the relevant physical regimes
and may even find (systematic or empirical) corrections, and then apply approximate theories to
treat larger systems.

In this paper, we use this strategy to examine the accuracy of a common approximation --- 
the noninteracting electron approximation for treating vibrationally coupled quantum 
transport. Approaches based on this approximation are sometimes used in combination with electronic structure 
theories to model nonequilibrium transport through a single molecular junction. For example, 
one may propagate the density matrix in a single electron basis with electronic-vibrational 
couplings, where the vibrations may be treated by the classical Ehrenfest approach or included 
as a self-energy correction.\cite{zha13:164121} Time-dependent density functional theory (TDDFT) in
combination with a classical treatment of the nuclear motion also belongs to this class
of approximations.\cite{Verdozzi06} In the absence of vibrational coupling this approach is exact for a tight-binding electronic Hamiltonian.  When the vibrational coupling is 
included, both electronic-vibrational and vibrationally induced correlation effects may 
participate in the real time quantum dynamics. To assess the errors introduced in this
approximation, we use the multilayer multiconfiguration time-dependent Hartree (ML-MCTDH) 
theory to describe the dynamics of the vibrational degrees of freedom exactly while maintaining a noninteracting electron 
approximation for the electronic dynamics. The correlation effects are analyzed in different
physical regimes by comparing with the the fully correlated simulation employing the ML-MCTDH-SQR theory. It is hoped that this study will provide
some insight into the commonly adopted noninteracting electron approximation.

The remainder of the paper is organized as follows. Section~\ref{modeltight} outlines 
the physical model and the observables of interest, and briefly discusses the simulation methods.
Section~\ref{results} presents numerical results for vibrationally coupled electron transport 
in different parameter regimes as well as comparisons with numerically exact simulations. Section~\ref{conclusions} concludes.

\section{Model and Simulation Methods}\label{modeltight}

\subsection{Model}

In this work we use a simple model for a molecular junction or a quantum dot to study correlation effects for vibrationally coupled 
electron transport.  The electronic part of the Hamiltonian is based on a tight-binding model,
where one electronic state of the molecular bridge is coupled to two electronic 
continua describing the left and the right electrodes. A distribution of harmonic oscillators 
is used to model the vibrational modes of the molecular bridge.  The total Hamiltonian is
given by
\begin{subequations}\label{Htot}
\begin{equation}
	\hat H = \hat H_{\rm el} + \hat H_{\rm nuc} + \hat H_{\rm el-nuc},
\end{equation}
where $\hat H_{\rm el}$, $\hat H_{\rm nuc}$, and $\hat H_{\rm el-nuc}$ 
describe the electronic, vibrational, and coupling terms, respectively
\begin{eqnarray}
	\hat H_{\rm el} &=& E_d d^+ d + \sum_{k_L} E_{k_L} c_{k_L}^+ c_{k_L}
	+ \sum_{k_R} E_{k_R} c_{k_R}^+ c_{k_R} \\
  &&	+ \sum_{k_L} V_{dk_L} ( d^+ c_{k_L} + c_{k_L}^+ d )
	+ \sum_{k_R} V_{dk_R} ( d^+ c_{k_R} + c_{k_R}^+ d ), \nonumber \\
\end{eqnarray}
\begin{equation}
	\hat H_{\rm nuc} = \frac{1}{2} \sum_j ( P_j^2 + \omega_j^2 Q_j^2 ), \\
\end{equation}
\begin{equation}
	\hat H_{\rm el-nuc} = d^+ d \sum_j \gamma_j Q_j.
\end{equation}
\end{subequations}
In the expression above $d^+/ d$, $c_{k_L}^+/ c_{k_L}$, $c_{k_R}^+/ c_{k_R}$ are the
fermionic creation/annihilation operators for the electronic states on the molecular 
bridge, the left and the right leads, respectively.  The corresponding electronic 
energies $E_{k_L}$, $E_{k_R}$ and the molecule-lead coupling strengths $V_{dk_L}$, $V_{dk_R}$, 
are defined through the energy-dependent level width functions
\begin{equation}
	\Gamma_L (E) = 2\pi \sum_{k_l} |V_{dk_L}|^2 \delta(E-E_{k_L}), \hspace{1cm}
	\Gamma_R (E) = 2\pi \sum_{k_r} |V_{dk_R}|^2 \delta(E-E_{k_R}).
\end{equation}
Employing a tight-binding model, the function $\Gamma (E)$ is given as
\begin{subequations}
\begin{equation}
        \Gamma (E) = \left\{ \begin{array}{ll} \frac{\alpha_e^2}{\beta_e^2} \sqrt{4\beta_e^2-E^2} 
		\hspace{1cm} & |E| \leq 2 |\beta_e| \\
                0 \hspace{1cm} &  |E| > 2 |\beta_e| \end{array}  \right.,
\end{equation}
\begin{equation}
	\Gamma_L (E) = \Gamma (E-\mu_L), \hspace{1cm}  \Gamma_R (E) = \Gamma (E-\mu_R),
\end{equation}
\end{subequations}
where $\beta_e$ and $\alpha_e$ are nearest-neighbor couplings between two lead sites 
and between the lead and the bridge state, respectively.  I.e., the width functions for 
the left and the right leads are obtained by shifting $\Gamma(E)$ relative to the chemical potentials 
of the corresponding leads.  We consider a case with two identical leads, in which the chemical 
potentials are given by 
\begin{equation}
	\mu_{L/R} = E_f \pm V/2,
\end{equation}
where $V$ is the bias voltage and $E_f$ the Fermi energy of the leads.  

Moreover, $P_j$ and $Q_j$ in Eq.\ (\ref{Htot}) denote the momentum and coordinate of the $j$th vibrational mode with frequency $\omega_j$.
The frequencies $\omega_j$ and electronic-vibrational coupling
constants $\gamma_j$ of the vibrational modes of the molecular junctions are modeled by a spectral density 
function\cite{leg87:1,Weiss93}
\begin{equation}
\label{discrete}
        J(\omega) = \frac{\pi} {2} \sum_{j} \frac{\gamma_{j}^{2}} {\omega_{j}}
        \delta(\omega - \omega_{j}).
\end{equation}
In this paper, the spectral density is chosen in  Ohmic form with an exponential cutoff
\begin{equation}
\label{ohmic}
        J_{\rm O}(\omega)  = \frac{\pi\lambda}{\omega_c} \omega e^{-\omega/\omega_c},
\end{equation}
where $\lambda$ is the reorganization energy.  
Both the electronic and the vibrational continua can be discretized using an appropriate 
scheme.\cite{wan03:1289}  In this paper, we employ 200-400 states
for each electronic lead, and a bath with 900 modes. In addition, we also consider the case of a single
vibrational mode.

The observable of interest in transport through molecular junctions
is the current for a given bias voltage, given by (in this paper we use atomic units where
$\hbar = e = 1$)
\begin{subequations}
\begin{equation}
	I_L(t) = - \frac{d N_L(t)} {dt} = -\frac{1}{{\rm tr}[\hat{\rho}]} {\rm tr}
        \left\{ \hat{\rho} e^{i\hat{H}t} i[\hat{H}, \hat{N}_{L}] e^{-i\hat{H}t} \right\},
\end{equation}
\begin{equation}
	I_R(t) = \frac{d N_R(t)} {dt} = \frac{1}{{\rm tr}[\hat{\rho}]} {\rm tr}
        \left\{ \hat{\rho} e^{i\hat{H}t} i[\hat{H}, \hat{N}_{R}] e^{-i\hat{H}t} \right\}.
\end{equation}
\end{subequations}
Here $\hat{N}_{\zeta} = \sum_{k_\zeta} c_{k_\zeta}^+ c_{k_\zeta}$
is the occupation number operator for the electrons in each lead ($\zeta=L, R$) and $\hat{\rho}$ 
is the initial density matrix representing a grand-canonical ensemble for each lead and a certain
occupation (occupied or unoccupied) for the bridge state
\begin{subequations}\label{Initden}
\begin{equation}
	\hat{\rho} = \hat{\rho}_d^0 \;{\rm exp} \left[ -\beta (\hat{H}_0 
          - \mu_L \hat{N}_L - \mu_R \hat{N}_R) \right],
\end{equation}
\begin{equation}
	\hat{H}_0 = \sum_{k_l} E_{k_l} c_{k_l}^+ c_{k_l}
	+ \sum_{k_r} E_{k_r} c_{k_r}^+ c_{k_r}  + \hat{H}_{\rm nuc}^0.
\end{equation}
\end{subequations}
That is, $\hat{\rho}_d^0$ is the initial reduced density matrix for the bridge state, which is 
chosen as a pure state representing an occupied or an empty bridge state, and 
$\hat{H}_{\rm nuc}^0$ defines the initial bath equilibrium distribution, e.g., $\hat{H}_{\rm nuc}$ 
given above in equilibrium with an empty bridge state or a shifted bath in equilibrium with 
an occupied bridge state. The dependence of the steady-state current on the initial density matrix  
has been discussed before.\cite{fer12:081412,wil13:045137}  In the context of the current work, it only affects the accuracy of the 
noninteracting electron approximation. 
To minimize the transient effects, the average current 
\begin{equation}
	I(t) = \frac{1}{2} [ I_R(t) + I_L(t) ],
\end{equation} 
will be used in the results presented below.

\subsection{Multilayer Multiconfiguration Time-Dependent Hartree Theory}

The physical observables are calculated by solving the time-dependent 
Schr\"{o}dinger equation employing the multilayer multiconfiguration time-dependent 
Hartree (ML-MCTDH) theory.\cite{wan03:1289,wan15:7951} Within the ML-MCTDH method the wave function 
$|\Psi (t) \rangle$ is expressed in a flexible, hierarchical form 
\begin{subequations}\label{psiml}
\begin{equation}
        |\Psi (t) \rangle = \sum_{j_1} \sum_{j_2} ... \sum_{j_p}
        A_{j_1j_2...j_p}(t) \prod_{\kappa=1}^{p}  |\varphi_{j_\kappa}^{(\kappa)} (t) \rangle,
\end{equation}
\begin{equation}
        |\varphi_{j_\kappa}^{(\kappa)}(t)\rangle =  \sum_{i_1} \sum_{i_2} ... \sum_{i_{Q(\kappa)}}
        B_{i_1i_2...i_{Q(\kappa)}}^{\kappa,j_\kappa}(t) \prod_{q=1}^{Q(\kappa)}  
	|v_{i_q}^{(\kappa,q)}(t) \rangle,
\end{equation}
\begin{equation}
        |v_{i_q}^{(\kappa,q)}(t)\rangle  = \sum_{\alpha_1} \sum_{\alpha_2} ... 
	\sum_{\alpha_{M(\kappa,q)}}
        C_{\alpha_1\alpha_2...\alpha_{M(\kappa,q)}}^{\kappa,q,i_q}(t) 
	\prod_{s=1}^{M(\kappa,q)}  
	|\xi_{\alpha_s}^{(\kappa,q,s)}(t) \rangle,
\end{equation}
\begin{equation}
	... \nonumber
\end{equation}
\end{subequations}
where $A_{j_1j_2...j_p}(t)$, $B_{i_1i_2...i_{Q(\kappa)}}^{\kappa,j_\kappa}(t)$, 
$C_{\alpha_1\alpha_2...\alpha_{M(\kappa,q)}}^{\kappa,q,i_q}(t)$, ..., are expansion coefficients of
the first (top) layer, second layer, third layer, and so on; and 
$|\varphi_{j_\kappa}^{(\kappa)} (t) \rangle$, $|v_{i_q}^{(\kappa,q)}(t) \rangle$, 
$|\xi_{\alpha_s}^{(\kappa,q,s)}(t) \rangle$, ..., are single particle functions (SPFs)
of the respective layers. The multilayer expansion is terminated at a particular level
by requiring the SPFs of the deepest layer to be time-independent, i.e., they are expanded
in static, primitive basis functions or contracted configurations within a few degrees of 
freedom.\cite{wan15:7951} SPFs of the second to last layer are then constructed using the 
expansion coefficients and the (static) SPFs of the last layer.  SPFs of all other layers 
are then built bottom-up according to Eq.~(\ref{psiml}). 

As in the underlying MCTDH method,\cite{Meyer90,Meyer09} the ML-MCTDH equations of motion\cite{wan03:1289,wan15:7951} are obtained by applying  the
Dirac-Frenkel variational principle. The implementation of the ML-MCTDH method follows a systematic 
streamlined procedure as described in detail previously.\cite{wan03:1289,wan15:7951}  On one hand, different parts of the Hamiltonian are built 
``bottom-up''. On the other hand, reduced density matrices needed in each 
layer are built ``top-down''. The matrices of mean-field operators is a combination 
of the two procedures.

The introduction of the recursive, dynamically optimized layering scheme in the ML-MCTDH 
wave function provides a great deal of flexibility in the trial wave function, which results 
in a tremendous gain in the ability to study large many-body quantum systems. 
This is demonstrated by many applications on simulating quantum dynamics of ultrafast 
electron transfer reactions in condensed phases.\cite{tho06:210,wan06:034114,kon06:1364,wan07:10369,kon07:11970,tho07:153313,cra07:144503,wan08:139,wan08:115005,ego08:214303,vel08:325,vel09:094109,wan10:78,zho12:581,wan12:22A504} 
The ML-MCTDH work of Manthe has introduced an even more adaptive formulation based on
a layered correlation discrete variable representation (CDVR).\cite{man08:164116,man09:054109} 
This important development potentially extends the applicability of ML-MCTDH theory 
to rather general systems described by a general form of the potential energy surface.

The original ML-MCTDH method was not directly applicable to systems of identical particles.  
This is because a Hartree product in the first quantized picture is only suitable to 
describe a configuration for a system of distinguishable particles.  To handle systems 
of identical particles explicitly, additional constraints need to be imposed since the 
exchange symmetry is not accounted for in the Schr\"odinger equation or the Dirac-Frenkel 
variational principle. To retain the multilayer form of the wave function, ML-MCTDH in
the second quantized form, the ML-MCTDH-SQR theory,\cite{wan09:024114} was proposed, where
the variation is carried out entirely in the abstract Fock space represented by the 
occupation number states. The ML-MCTDH-SQR theory has seen several promising 
applications.\cite{wan11:244506,fer12:081412,wan13:134704,wan13:7431,wil13:045137}

\subsection{Noninteracting Electron Approximation}

ML-MCTDH-SQR simulations taking full account of electron-electron and electronic-vibrational correlations 
can be computationally demanding. Thus, it is of interest to seek less demanding approximate solutions. 
One approximation is to adopt a noninteracting electron picture, that is, neglecting 
electron-electron correlation effects.
To formulate a noninteracting electron theory of vibrationally coupled electron transport, we consider the single-electron Hamiltonian underlying the many-electron Hamiltonian given in Eq.\ (\ref{Htot}),
\begin{eqnarray}
	\hat h &=& E_d |d\rangle\langle d|  + \sum_{k_L} E_{k_L} |k_L\rangle\langle k_L|
	+ \sum_{k_R} E_{k_R} |k_R\rangle\langle k_R| \\
  &&	+ \sum_{k_L} V_{dk_L} ( |d\rangle\langle k_L| + |k_L\rangle\langle d| )
	+ \sum_{k_R} V_{dk_R} ( |d\rangle\langle k_R| + |k_R\rangle\langle d| ) , \nonumber \\
	&& + \frac{1}{2} \sum_j ( P_j^2 + \omega_j^2 Q_j^2 )
	+|d\rangle\langle d| \sum_j \gamma_j Q_j,
	\label{Hsp}
\end{eqnarray}
where $|k_{L/R}\rangle$, $|d\rangle$ denote the electronic single particle states in the left/right leads and at the molecule bridge, respectively. The solution of the time-dependent Schr\"odinger equation for the single-electron Hamiltonian (\ref{Hsp}) represents still a many-body problem, due to electronic-vibrational coupling. To solve it, we use also the ML-MCTDH method, similar as in our previous work on electron transfer at dye-semiconductor interfaces.\cite{Thoss04b,Li15}

To calculate transport properties using the noninteracting electron approximation, the time-dependent Schr\"odinger equation is solved for a set of initial states. The approximation to the single electron density matrix is then obtained by summing over these wave functions weighted according to their initial occupations
\begin{eqnarray}\label{dm}
\hat \rho_{se} (t) & = & 
\sum_j p(j) |\psi_j(t)\rangle\langle \psi_j(t)|,
\end{eqnarray}
where $p(j)$ denotes the initial occupation determined by the distribution in Eq.\ (\ref{Initden}). 
The initial wave function is given by
\begin{eqnarray}
|\psi_j(0)\rangle &=&
|k_j\rangle|v_0\rangle
\end{eqnarray}
if lead state $|k_j\rangle$ is initially occupied
or 
\begin{eqnarray}
|\psi_j(0)\rangle &=&
|d\rangle|v_0\rangle
\end{eqnarray}
if the electronic bridge state is initially occupied. Furthermore, $|v_0\rangle$ denotes the initial vibrational state, which in all results presented below is the ground vibrational state of the occupied or unoccupied molecular bridge.
Based on the single electron density matrix (\ref{dm}), the current within the noninteracting electron approximation is given by
\begin{eqnarray}
	I_L(t) = - \frac{d N_L(t)} {dt} = 
	-\frac{d} {dt}\sum_{k_L}{\rm tr}
        \left\{ |k_L\rangle\langle k_L|\, \hat\rho_{se}(t)\right\},
\end{eqnarray}
and similar for $I_R(t)$.

It is noted that the noninteracting electron approximation introduced above is
exact for vanishing electronic-vibrational coupling, i.e.\ for the noninteracting transport problem.
In this work, we examine the accuracy of such an approximation in the presence of 
electronic-vibrational coupling. It is also noted that the noninteracting electron approximation to vibrationally coupled electron transport is similar to the inelastic scattering theory approach to that problem.\cite{Cizek04,Ness05,Benesch06,Benesch08} Both approaches treat the transport of independent electrons coupled to the vibrational degrees of freedom. In contrast to the scattering theory approach, the ML-MCTDH treatment of the noninteracting electron approximation is not limited to a few vibrational modes.

\section{Results and Discussion}\label{results}

We will assess the accuracy of the noninteracting 
electron transport approximation by comparing the results from this approach with those
obtained from the fully converged, numerically exact ML-MCTDH-SQR theory.  In both simulations the
vibrational degrees of freedom are treated via the converged ML-MCTDH approach.  The difference is that in the ML-MCTDH-SQR calculations the (vibrationally induced) electron-electron correlations are fully accounted for whereas
the noninteracting electron approximation lacks such a treatment.  To distinguish the two approaches, we call ML-MCTDH-SQR calculations the ``full'' simulation.
In all results presented below, the temperature is $T=0$ and the tight-binding parameters for the function $\Gamma (E)$ are $\alpha_e = 0.2$ eV, 
$\beta_e = 1$ eV, corresponding to a moderate molecule-lead coupling and a bandwidth of 4 eV. 

\begin{figure}[!ht]
\begin{flushleft}
(a)
\end{flushleft}

\includegraphics[clip,width=0.5\textwidth]{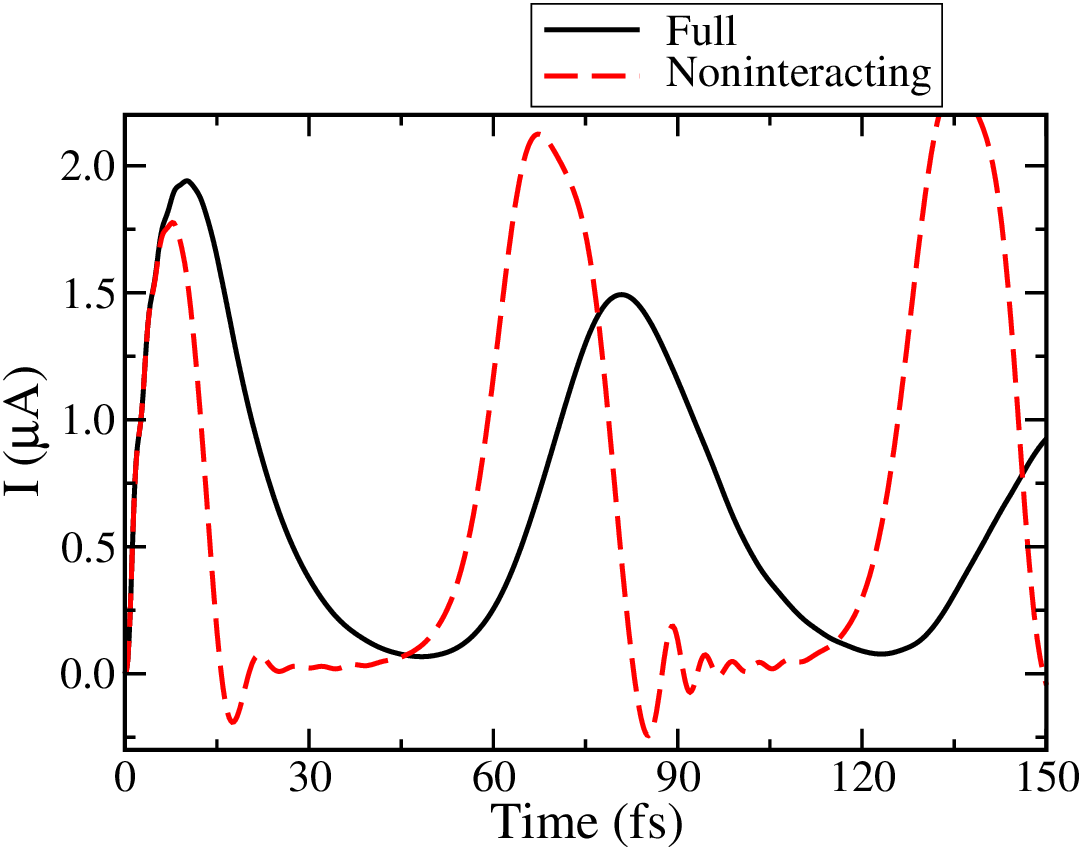}

\begin{flushleft}
(b)
\end{flushleft}
\includegraphics[clip,width=0.5\textwidth]{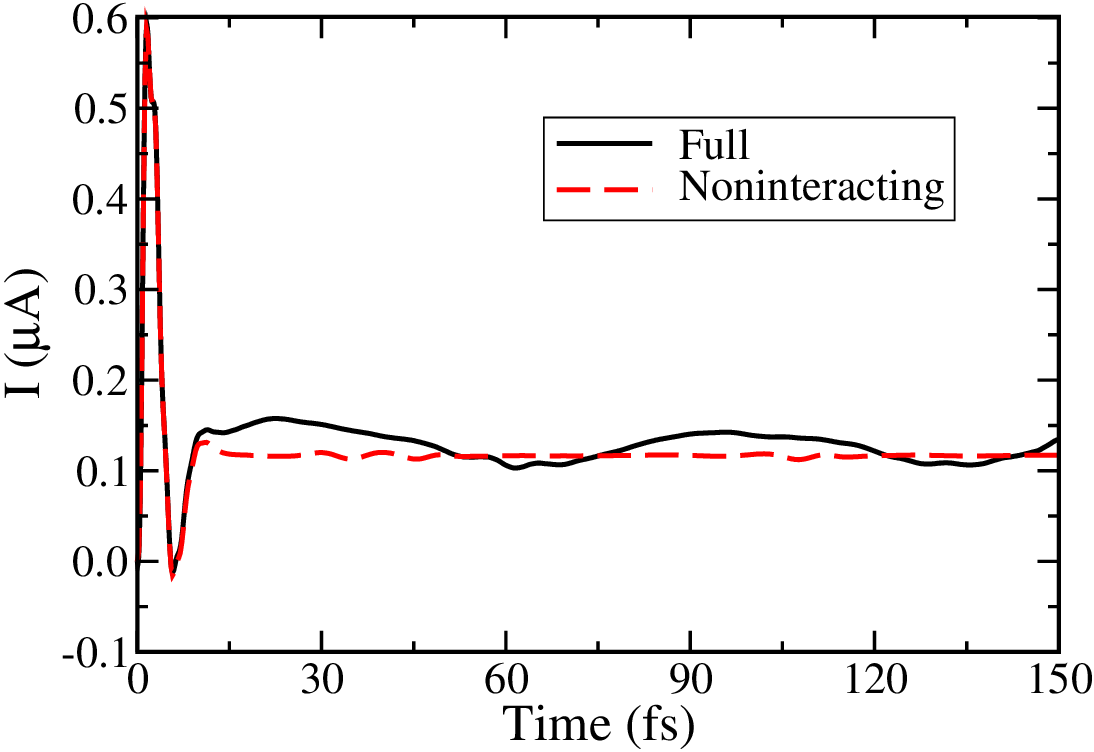}

\caption{\label{fig1} Comparison of the time-dependent current $I(t)$ between the noninteracting
electron approximation and the full ML-MCTDH-SQR simulation where a single vibrational mode
is coupled to the bridge state. The frequency is $\omega = 500$ cm$^{-1}$ and the reorganization 
energy is $\lambda = 2000$ cm$^{-1}$. The bias voltage is $V = 0.1$V and the initial condition 
is: (a) an occupied bridge state with the mode's coordinate shifted to be in equilibrium with it; 
(b) an empty bridge state with an unshifted mode.  }
\end{figure}

We first consider a model, where the discrete state $E_d$ is located 0.5 eV above the Fermi energy of the leads $E_f$.  
Figure~\ref{fig1}a shows the time-dependent current for the case with a single vibrational mode.  Initially, the bridge state is occupied and the vibrational mode is in equilibrium
with the occupied bridge state. Significant oscillations of the current on the time scale of the vibrational mode are observed in the transient 
current for this initial condition, which will be quenched for long times. Compared 
with the full ML-MCTDH-SQR simulation, the noninteracting electron approximation reproduces 
$I(t)$ only for very short time. Since it does not include vibrational nonequilibrium  effect induced by 
electron transport, it incorrectly predicts an increase in the amplitude of initial oscillation 
whereas the full simulation predicts a damped oscillation. The noninteracting calculations also exhibits spurious fast oscillations of the current. 

Figure~\ref{fig1}b shows the time-dependent current for the same set of parameters
but with a different initial state: an unoccupied bridge state and an unshifted vibrational mode.
Within the same time scale as in Figure~\ref{fig1}a, the noninteracting electron approximation provides a much better agreement 
with the full ML-MCTDH-SQR simulation result. Although it exaggerates the decoherence of vibrational oscillations the 
current, it reproduces the first short time transient oscillation, which is of electronic origin, and predicts a steady-state 
current that agrees approximately with the average of the full simulation result.  

\begin{figure}[!h]

\includegraphics[clip,width=0.5\textwidth]{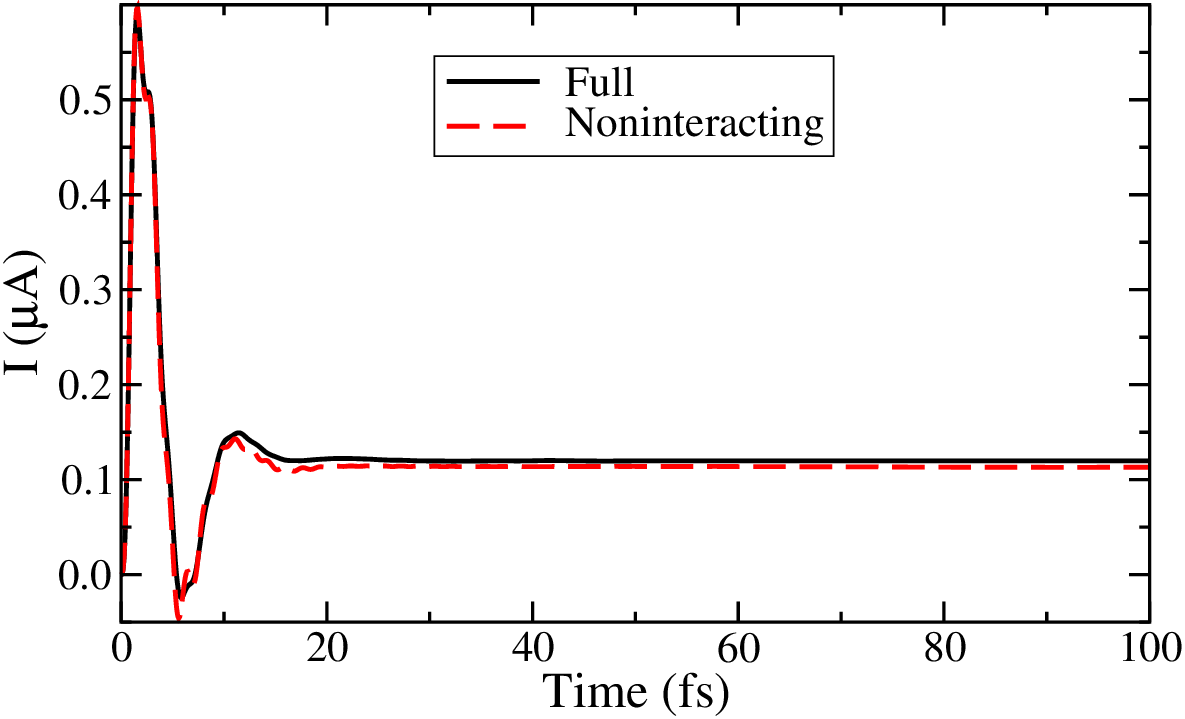}

\caption{\label{fig2} Same as Fig.~\ref{fig1}b but for a vibrational bath modeled by an Ohmic spectral 
density. The characteristic frequency is $\omega_c = 500$ cm$^{-1}$ and the reorganization 
energy is $\lambda = 2000$ cm$^{-1}$. The initial condition is specified by an empty bridge state 
with an unshifted vibrational bath.   }
\end{figure}

This observation suggests that the accuracy of the noninteracting electron approximation depends
on the initial condition. If the initial density matrix is closer to the steady state, then
the effect of vibrationally induced electron correlation is smaller, which renders the noninteracting electron 
approximation more accurate.  In the example above, an initially unoccupied bridge state with an unshifted
vibrational mode is closer to the steady state distribution. Thus, the noninteracting electron approximation
for this initial condition agrees better with the full ML-MCTDH-SQR simulation.  One would expect
that if the single vibrational mode is replaced by a vibrational bath, the electronic coherence will be
quenched more efficiently such that the agreement between the noninteracting electron approximation and the
full ML-MCTDH-SQR simulation would improve. This is indeed the case, as shown in 
Figure~\ref{fig2}.

\begin{figure}[!ht]
\begin{flushleft}
(a)
\end{flushleft}
\includegraphics[clip,width=0.5\textwidth]{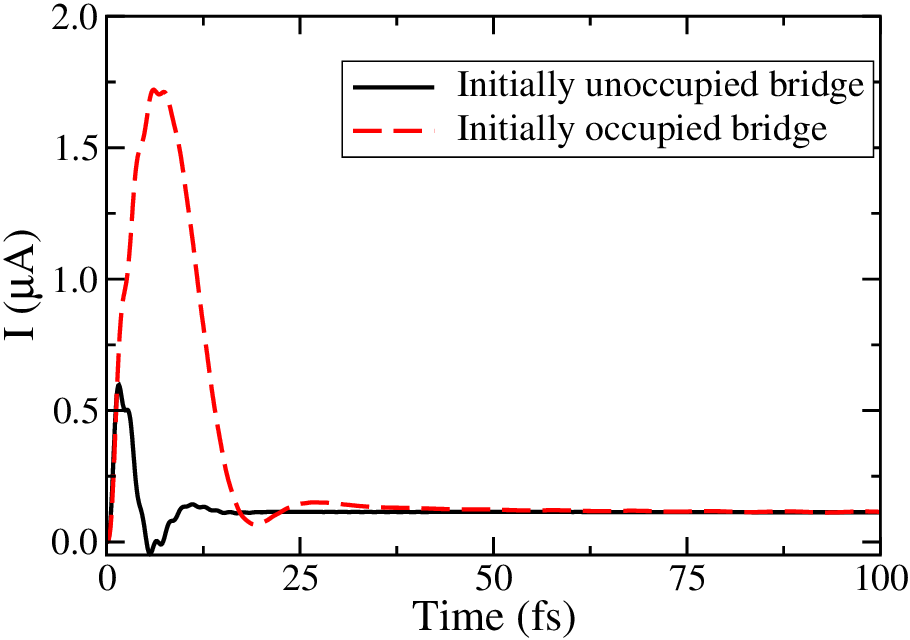}

\begin{flushleft}
(b)

\end{flushleft}
\includegraphics[clip,width=0.5\textwidth]{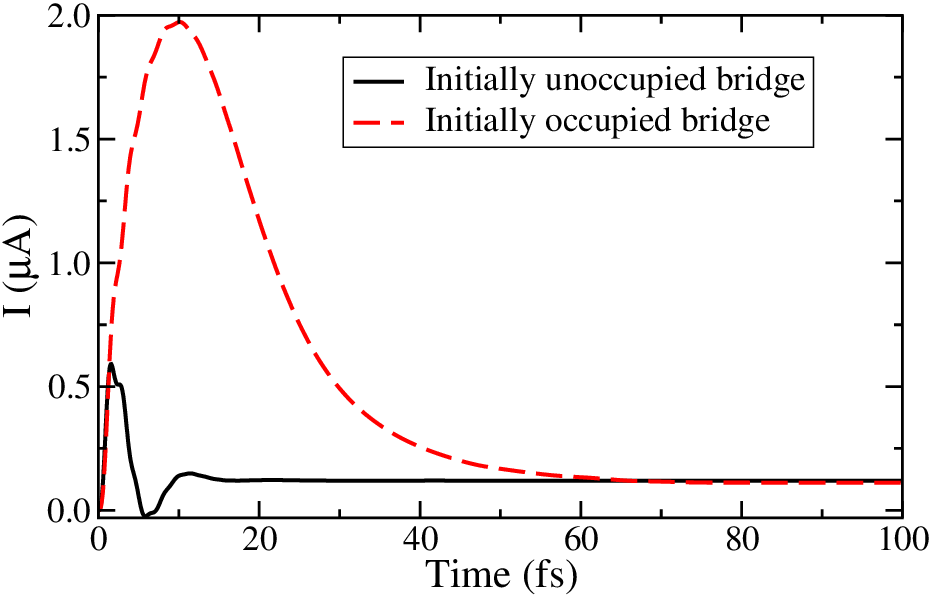}

\caption{\label{fig3} Time-dependent current at different initial conditions: (a) 
noninteracting electron approximation, (b) full ML-MCTDH-SQR simulation. The characteristic 
frequency for the vibrational bath is $\omega_c = 500$ cm$^{-1}$ and the reorganization energy is 
$\lambda = 2000$ cm$^{-1}$. The bias voltage is $V = 0.1$V. }
\end{figure}

If different initial conditions give the same steady state current within a reasonably short time, 
one may argue that although the noninteracting electron approximation does not give the correct
transient dynamics, it may still predict the correct stationary current. This is to some extent correct,
as shown in Figure~\ref{fig3}, where the parameters are the same as in Figure~\ref{fig2}.  Results for two 
initial conditions are plotted corresponding to an occupied or an unoccupied bridge state. In each case the vibrational bath 
is in equilibrium with the bridge state.  It is seen that the two initial conditions give the same 
stationary current within the simulation timescale.  The stationary current from the noninteracting 
electron approximation, as shown in Figure~\ref{fig2}, agrees with that from the full ML-MCTDH-SQR 
simulation.

\begin{figure}[!ht]
\begin{flushleft}
(a)
\end{flushleft}
\includegraphics[clip,width=0.5\textwidth]{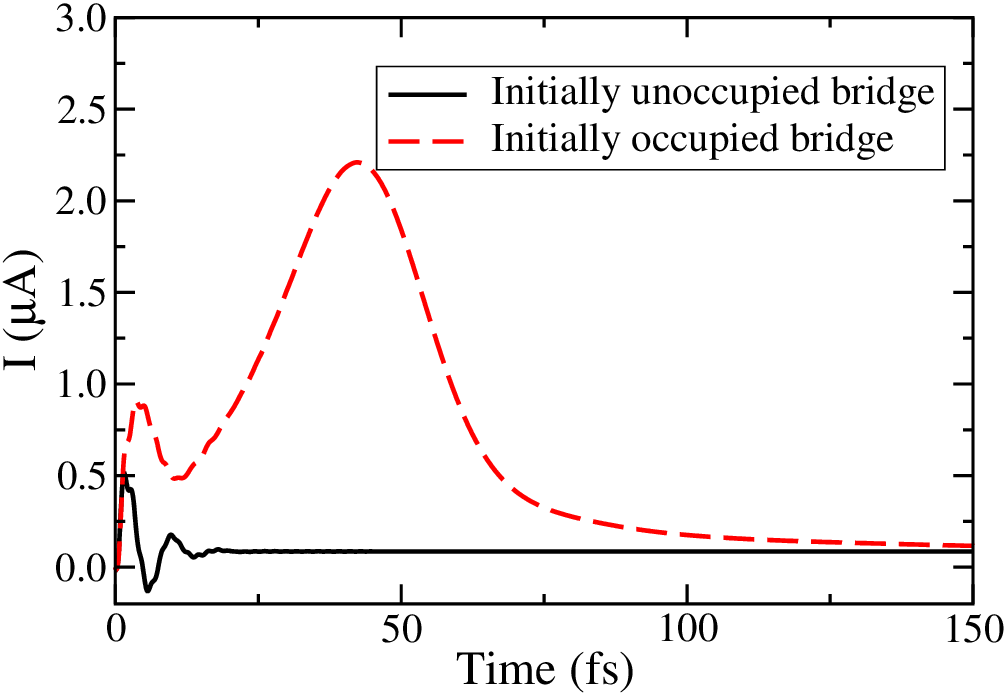}

\begin{flushleft}
(b)

\end{flushleft}
\includegraphics[clip,width=0.5\textwidth]{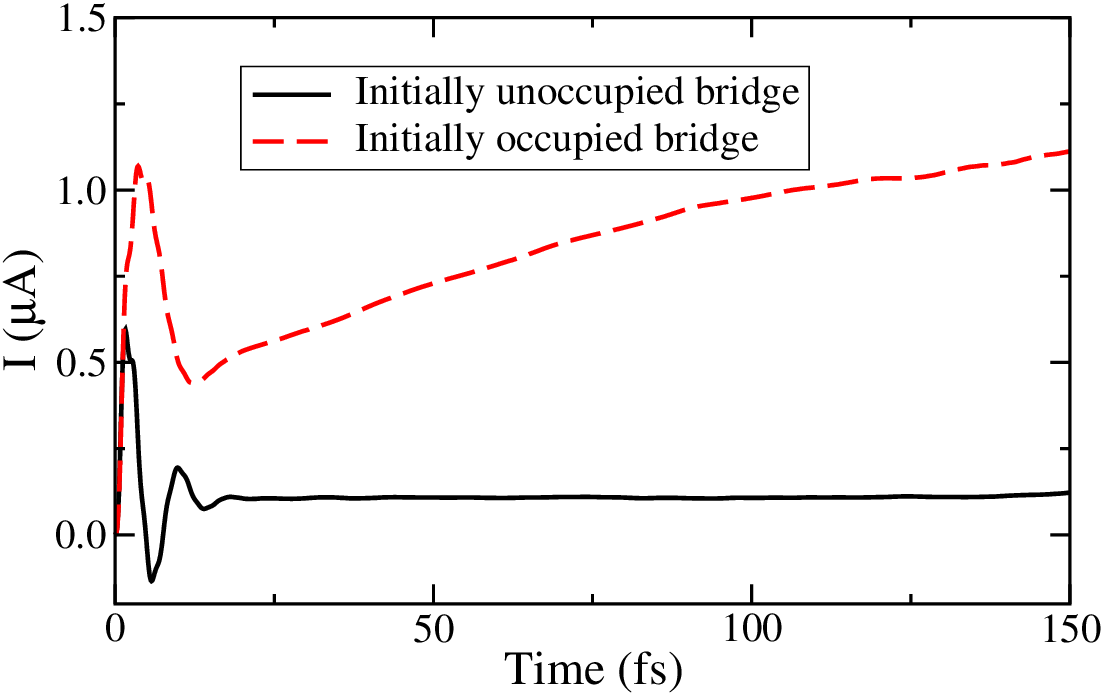}

\caption{\label{fig4} Time-dependent current at different initial conditions: (a) 
noninteracting electron approximation, (b) full ML-MCTDH-SQR simulation. The characteristic 
frequency for the vibrational bath is $\omega_c = 100$~cm$^{-1}$ and the reorganization energy is 
$\lambda = 3000$~cm$^{-1}$. The bias voltage is $V = 0.1$~V. }
\end{figure}

\begin{figure}[!h]

\includegraphics[clip,width=0.5\textwidth]{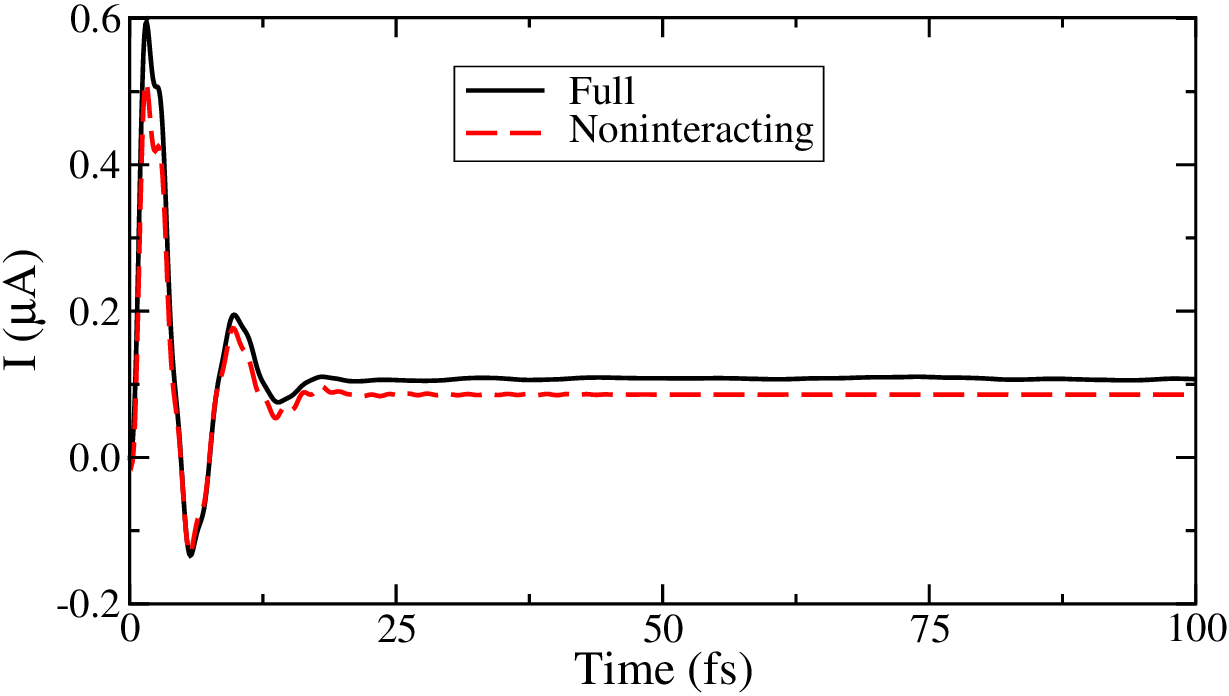}

\caption{\label{fig5} Comparison of the stationary current between the noninteracting electron 
approximation and the full ML-MCTDH-SQR simulation for the initially unoccupied bridge state and
an unshifted vibrational bath.  The parameters are the same as in Fig.~\ref{fig4}.  }
\end{figure}

As discussed previously,\cite{wil13:045137}, for certain parameter regime, in particular small bias voltage, low bath characteristic frequency of the vibrational bath and strong electronic-vibrational coupling, the bridge state population and the time-dependent current may exhibit long-time bistability behavior.  Figure~\ref{fig4} is an example of this phenomenon.
It is seen that the noninteracting electron approximation incorrectly predicts that the two 
different initial conditions lead to the same stationary current within a very short time.
Comparing with the full ML-MCTDH-SQR simulation it can be concluded that the bistability
behavior is due to vibrationally induced correlation, which cannot be captured
by the noninteracting electron approximation. Interestingly, if one picks the ``correct''
initial condition based on physical intuition (in this case an unoccupied bridge state and
an unshifted vibrational bath), then the stationary current from the noninteracting electron 
approximation does not deviate much from the full ML-MCTDH-SQR value. As shown in 
Figure~\ref{fig5}, the error of the noninteracting electron approximation is only 20\% for
this set of parameters.

\begin{figure}[!ht]
\begin{flushleft}
(a)
\end{flushleft}
\includegraphics[clip,width=0.5\textwidth]{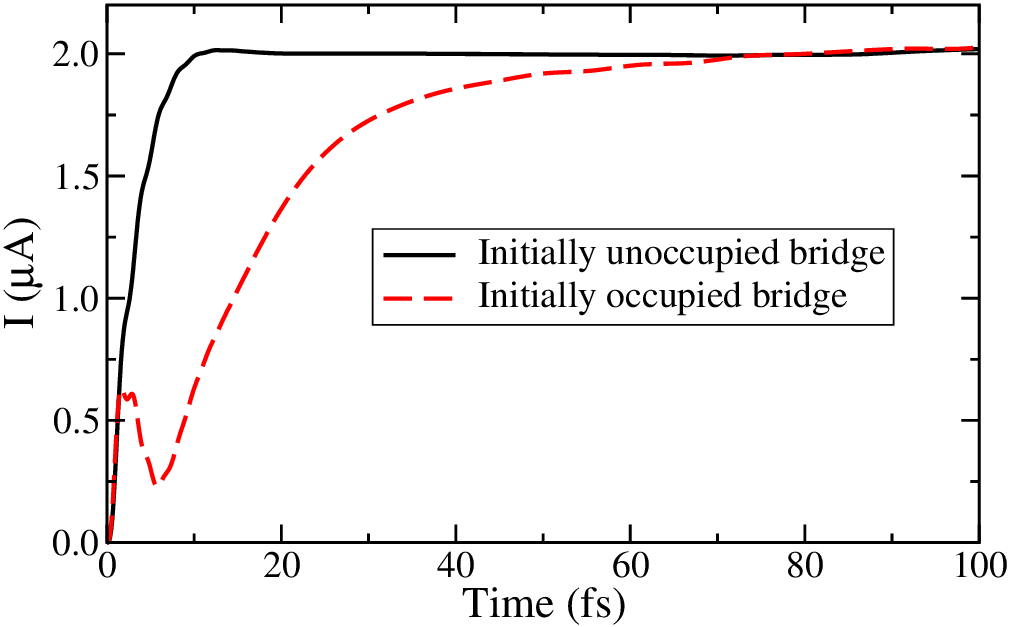}

\begin{flushleft}
(b)

\end{flushleft}
\includegraphics[clip,width=0.5\textwidth]{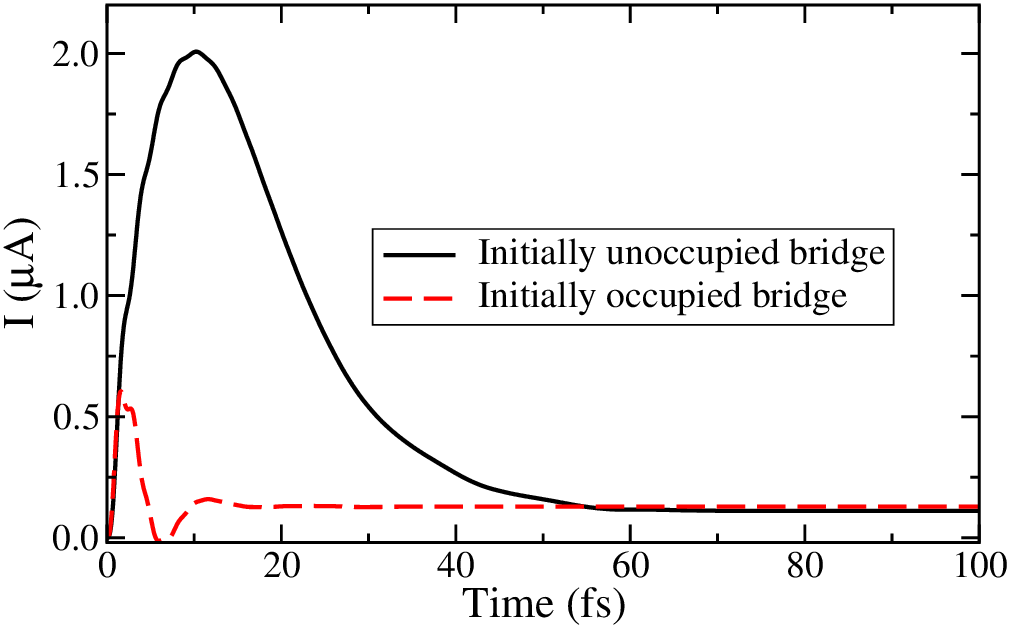}

\caption{\label{fig6} Time-dependent current at different initial conditions: (a) 
noninteracting electron approximation, (b) full ML-MCTDH-SQR simulation. The bridge
state has the same energy as the Fermi level, $E_d - E_f=0$. The characteristic 
frequency for the vibrational bath is $\omega_c = 500$~cm$^{-1}$ and the reorganization energy is 
$\lambda = 2000$~cm$^{-1}$. The bias voltage is $V = 0.1$~V. }
\end{figure}

The most severe failure of the noninteracting electron approximation occurs when the vibrationally 
induced correlation effect becomes dominant.  One such example is the regime of phonon blockade, as shown
in Figure~\ref{fig6}. The electronic parameters are the same as above except that the energy 
of the discrete state $E_d$ coincides with the Fermi energy of the leads $E_f$. The usual 
qualitative interpretation of the observed suppression of the current due to phonon blockade is that the polaron shift brings the bridge
state out of the bias window. Figure~\ref{fig6} shows that this is due to vibrationally 
induced correlation, because the noninteracting electron approximation predicts an incorrect value of the
stationary current.

\section{Concluding Remarks}\label{conclusions}

In this paper, we have assessed the validity of a the noninteracting electron approximation to describe transient and steady state transport in models of molecular junctions with electronic-vibrational interaction.  Within the noninteracting electron approximation, a single electron description is adopted but the interaction with the vibrational degrees of freedom is still described completely using the ML-MCTDH method. The assessment is based on a comparison with numerically exact results for the interacting transport problem obtained with the ML-MCTDH-SQR method.

The results show that the noninteracting electron approximation provides a good representation of the short time dynamics, but may fail to describe the longer time dynamics and the steady state current. This is particularly the case for parameter regimes that involve significant vibrationally induced correlation effects, such as, e.g., in the phonon blockade regime.
The validity of the noninteracting electron approximation can be improved by using an initial state that is close to the steady state.

\section*{Acknowledgments}
This work has been supported by the National Science Foundation CHE-1500285 (HW)
and the Deutsche Forschungsgemeinschaft (DFG) (MT), and used resources
of the National Energy Research Scientific Computing Center, a DOE Office of Science 
User Facility supported by the Office of Science of the U.S. Department of Energy under 
Contract No. DE-AC02-05CH11231.

\pagebreak

%
%


\clearpage

\clearpage
~
\vspace{3cm}

\clearpage

\clearpage

\clearpage
~
\vspace{3cm}

\clearpage

\begin{thebibliography}{100}
\expandafter\ifx\csname bibnamefont\endcsname\relax
  \def\bibnamefont#1{#1}\fi
\expandafter\ifx\csname bibfnamefont\endcsname\relax
  \def\bibfnamefont#1{#1}\fi
\expandafter\ifx\csname url\endcsname\relax
  \def\url#1{\texttt{#1}}\fi
\expandafter\ifx\csname urlprefix\endcsname\relax\def\urlprefix{URL }\fi
\providecommand{\bibinfo}[2]{#2}
\providecommand{\eprint}[2][]{\url{#2}}

\bibitem{ree97:252}
\bibinfo{author}{\bibfnamefont{M.}~\bibnamefont{Reed}},
  \bibinfo{author}{\bibfnamefont{C.}~\bibnamefont{Zhou}},
  \bibinfo{author}{\bibfnamefont{C.}~\bibnamefont{Muller}},
  \bibinfo{author}{\bibfnamefont{T.}~\bibnamefont{Burgin}}, \bibnamefont{and}
  \bibinfo{author}{\bibfnamefont{J.}~\bibnamefont{Tour}},
  \bibinfo{journal}{Science} \textbf{\bibinfo{volume}{278}},
  \bibinfo{pages}{252} (\bibinfo{year}{1997}).

\bibitem{joa00:541}
\bibinfo{author}{\bibfnamefont{C.}~\bibnamefont{Joachim}},
  \bibinfo{author}{\bibfnamefont{J.}~\bibnamefont{Gimzewski}},
  \bibnamefont{and} \bibinfo{author}{\bibfnamefont{A.}~\bibnamefont{Aviram}},
  \bibinfo{journal}{Nature (London)} \textbf{\bibinfo{volume}{408}},
  \bibinfo{pages}{541} (\bibinfo{year}{2000}).

\bibitem{Nitzan01}
\bibinfo{author}{\bibfnamefont{A.}~\bibnamefont{Nitzan}},
  \bibinfo{journal}{Annu. Rev. Phys. Chem.} \textbf{\bibinfo{volume}{52}},
  \bibinfo{pages}{681} (\bibinfo{year}{2001}).

\bibitem{nit03:1384}
\bibinfo{author}{\bibfnamefont{A.}~\bibnamefont{Nitzan}} \bibnamefont{and}
  \bibinfo{author}{\bibfnamefont{M.~A.} \bibnamefont{Ratner}},
  \bibinfo{journal}{Science} \textbf{\bibinfo{volume}{300}},
  \bibinfo{pages}{1384} (\bibinfo{year}{2003}).

\bibitem{Cuniberti05}
\bibinfo{author}{\bibfnamefont{G.}~\bibnamefont{Cuniberti}},
  \bibinfo{author}{\bibfnamefont{G.}~\bibnamefont{Fagas}}, \bibnamefont{and}
  \bibinfo{author}{\bibfnamefont{K.}~\bibnamefont{Richter}},
  \emph{\bibinfo{title}{Introducing Molecular Electronics}}
  (\bibinfo{publisher}{Springer}, \bibinfo{address}{Heidelberg},
  \bibinfo{year}{2005}).

\bibitem{Selzer06}
\bibinfo{author}{\bibfnamefont{Y.}~\bibnamefont{Selzer}} \bibnamefont{and}
  \bibinfo{author}{\bibfnamefont{D.~L.} \bibnamefont{Allara}},
  \bibinfo{journal}{Annu. Rev. Phys. Chem.} \textbf{\bibinfo{volume}{57}},
  \bibinfo{pages}{593} (\bibinfo{year}{2006}).

\bibitem{Venkataraman06}
\bibinfo{author}{\bibfnamefont{L.}~\bibnamefont{Venkataraman}},
  \bibinfo{author}{\bibfnamefont{J.~E.} \bibnamefont{Klare}},
  \bibinfo{author}{\bibfnamefont{C.}~\bibnamefont{Nuckolls}},
  \bibinfo{author}{\bibfnamefont{M.~S.} \bibnamefont{Hybertsen}},
  \bibnamefont{and} \bibinfo{author}{\bibfnamefont{M.~L.}
  \bibnamefont{Steigerwald}}, \bibinfo{journal}{Nature}
  \textbf{\bibinfo{volume}{442}}, \bibinfo{pages}{904} (\bibinfo{year}{2006}).

\bibitem{Chen07}
\bibinfo{author}{\bibfnamefont{F.}~\bibnamefont{Chen}},
  \bibinfo{author}{\bibfnamefont{J.}~\bibnamefont{Hihath}},
  \bibinfo{author}{\bibfnamefont{Z.}~\bibnamefont{Huang}},
  \bibinfo{author}{\bibfnamefont{X.}~\bibnamefont{Li}}, \bibnamefont{and}
  \bibinfo{author}{\bibfnamefont{N.}~\bibnamefont{Tao}},
  \bibinfo{journal}{Annu. Rev. Phys. Chem.} \textbf{\bibinfo{volume}{58}},
  \bibinfo{pages}{535} (\bibinfo{year}{2007}).

\bibitem{Galperin08b}
\bibinfo{author}{\bibfnamefont{M.}~\bibnamefont{Galperin}},
  \bibinfo{author}{\bibfnamefont{M.~A.} \bibnamefont{Ratner}},
  \bibinfo{author}{\bibfnamefont{A.}~\bibnamefont{Nitzan}}, \bibnamefont{and}
  \bibinfo{author}{\bibfnamefont{A.}~\bibnamefont{Troisi}},
  \bibinfo{journal}{Science} \textbf{\bibinfo{volume}{319}},
  \bibinfo{pages}{1056} (\bibinfo{year}{2008}).

\bibitem{Cuevas10}
\bibinfo{author}{\bibfnamefont{J.}~\bibnamefont{Cuevas}} \bibnamefont{and}
  \bibinfo{author}{\bibfnamefont{E.}~\bibnamefont{Scheer}},
  \emph{\bibinfo{title}{Molecular Electronics: An Introduction to Theory and
  Experiment}} (\bibinfo{publisher}{World Scientific},
  \bibinfo{address}{Singapore}, \bibinfo{year}{2010}).

\bibitem{par00:57}
\bibinfo{author}{\bibfnamefont{H.}~\bibnamefont{Park}},
  \bibinfo{author}{\bibfnamefont{J.}~\bibnamefont{Park}},
  \bibinfo{author}{\bibfnamefont{A.}~\bibnamefont{Lim}},
  \bibinfo{author}{\bibfnamefont{E.}~\bibnamefont{Anderson}},
  \bibinfo{author}{\bibfnamefont{A.}~\bibnamefont{Alivisatos}},
  \bibnamefont{and} \bibinfo{author}{\bibfnamefont{P.}~\bibnamefont{McEuen}},
  \bibinfo{journal}{Nature (London)} \textbf{\bibinfo{volume}{407}},
  \bibinfo{pages}{57} (\bibinfo{year}{2000}).

\bibitem{cui01:571}
\bibinfo{author}{\bibfnamefont{X.}~\bibnamefont{Cui}},
  \bibinfo{author}{\bibfnamefont{A.}~\bibnamefont{Primak}},
  \bibinfo{author}{\bibfnamefont{X.}~\bibnamefont{Zarate}},
  \bibinfo{author}{\bibfnamefont{J.}~\bibnamefont{Tomfohr}},
  \bibinfo{author}{\bibfnamefont{O.}~\bibnamefont{Sankey}},
  \bibinfo{author}{\bibfnamefont{A.}~\bibnamefont{Moore}},
  \bibinfo{author}{\bibfnamefont{T.}~\bibnamefont{Moore}},
  \bibinfo{author}{\bibfnamefont{D.}~\bibnamefont{Gust}},
  \bibinfo{author}{\bibfnamefont{G.}~\bibnamefont{Harris}}, \bibnamefont{and}
  \bibinfo{author}{\bibfnamefont{S.}~\bibnamefont{Lindsay}},
  \bibinfo{journal}{Science} \textbf{\bibinfo{volume}{294}},
  \bibinfo{pages}{571} (\bibinfo{year}{2001}).

\bibitem{par02:722}
\bibinfo{author}{\bibfnamefont{J.}~\bibnamefont{Park}},
  \bibinfo{author}{\bibfnamefont{A.}~\bibnamefont{Pasupathy}},
  \bibinfo{author}{\bibfnamefont{J.}~\bibnamefont{Goldsmith}},
  \bibinfo{author}{\bibfnamefont{C.}~\bibnamefont{Chang}},
  \bibinfo{author}{\bibfnamefont{Y.}~\bibnamefont{Yaish}},
  \bibinfo{author}{\bibfnamefont{J.}~\bibnamefont{Petta}},
  \bibinfo{author}{\bibfnamefont{M.}~\bibnamefont{Rinkoski}},
  \bibinfo{author}{\bibfnamefont{J.}~\bibnamefont{Sethna}},
  \bibinfo{author}{\bibfnamefont{H.}~\bibnamefont{Abruna}},
  \bibinfo{author}{\bibfnamefont{P.}~\bibnamefont{McEuen}}, \bibnamefont{and}
  \bibinfo{author}{\bibfnamefont{D.}~\bibnamefont{Ralph}},
  \bibinfo{journal}{Nature (London)} \textbf{\bibinfo{volume}{417}},
  \bibinfo{pages}{722} (\bibinfo{year}{2002}).

\bibitem{smi02:906}
\bibinfo{author}{\bibfnamefont{R.}~\bibnamefont{Smit}},
  \bibinfo{author}{\bibfnamefont{Y.}~\bibnamefont{Noat}},
  \bibinfo{author}{\bibfnamefont{C.}~\bibnamefont{Untiedt}},
  \bibinfo{author}{\bibfnamefont{N.}~\bibnamefont{Lang}},
  \bibinfo{author}{\bibfnamefont{M.}~\bibnamefont{van Hemert}},
  \bibnamefont{and} \bibinfo{author}{\bibfnamefont{J.}~\bibnamefont{van
  Ruitenbeek}}, \bibinfo{journal}{Nature (London)}
  \textbf{\bibinfo{volume}{419}}, \bibinfo{pages}{906} (\bibinfo{year}{2002}).

\bibitem{rei02:176804}
\bibinfo{author}{\bibfnamefont{J.}~\bibnamefont{Reichert}},
  \bibinfo{author}{\bibfnamefont{R.}~\bibnamefont{Ochs}},
  \bibinfo{author}{\bibfnamefont{D.}~\bibnamefont{Beckmann}},
  \bibinfo{author}{\bibfnamefont{H.}~\bibnamefont{Weber}},
  \bibinfo{author}{\bibfnamefont{M.}~\bibnamefont{Mayor}}, \bibnamefont{and}
  \bibinfo{author}{\bibfnamefont{H.}~\bibnamefont{von Lohneysen}},
  \bibinfo{journal}{Phys. Rev. Lett.} \textbf{\bibinfo{volume}{88}},
  \bibinfo{pages}{176804} (\bibinfo{year}{2002}).

\bibitem{xu03:1221}
\bibinfo{author}{\bibfnamefont{B.}~\bibnamefont{Xu}} \bibnamefont{and}
  \bibinfo{author}{\bibfnamefont{N.}~\bibnamefont{Tao}},
  \bibinfo{journal}{Science} \textbf{\bibinfo{volume}{301}},
  \bibinfo{pages}{1221} (\bibinfo{year}{2003}).

\bibitem{qiu04:206102}
\bibinfo{author}{\bibfnamefont{X.}~\bibnamefont{Qiu}},
  \bibinfo{author}{\bibfnamefont{G.}~\bibnamefont{Nazin}}, \bibnamefont{and}
  \bibinfo{author}{\bibfnamefont{W.}~\bibnamefont{Ho}}, \bibinfo{journal}{Phys.
  Rev. Lett.} \textbf{\bibinfo{volume}{92}}, \bibinfo{pages}{206102}
  (\bibinfo{year}{2004}).

\bibitem{liu04:11371}
\bibinfo{author}{\bibfnamefont{N.}~\bibnamefont{Liu}},
  \bibinfo{author}{\bibfnamefont{N.}~\bibnamefont{Pradhan}}, \bibnamefont{and}
  \bibinfo{author}{\bibfnamefont{W.}~\bibnamefont{Ho}}, \bibinfo{journal}{J.
  Chem. Phys.} \textbf{\bibinfo{volume}{120}}, \bibinfo{pages}{11371}
  (\bibinfo{year}{2004}).

\bibitem{elb05:8815}
\bibinfo{author}{\bibfnamefont{M.}~\bibnamefont{Elbing}},
  \bibinfo{author}{\bibfnamefont{R.}~\bibnamefont{Ochs}},
  \bibinfo{author}{\bibfnamefont{M.}~\bibnamefont{Koentopp}},
  \bibinfo{author}{\bibfnamefont{M.}~\bibnamefont{Fischer}},
  \bibinfo{author}{\bibfnamefont{C.}~\bibnamefont{von {H\"anisch}}},
  \bibinfo{author}{\bibfnamefont{F.}~\bibnamefont{Weigend}},
  \bibinfo{author}{\bibfnamefont{F.}~\bibnamefont{Evers}},
  \bibinfo{author}{\bibfnamefont{H.}~\bibnamefont{Weber}}, \bibnamefont{and}
  \bibinfo{author}{\bibfnamefont{M.}~\bibnamefont{Mayor}},
  \bibinfo{journal}{Proc. Natl. Acad. Sci. USA} \textbf{\bibinfo{volume}{102}},
  \bibinfo{pages}{8815} (\bibinfo{year}{2005}).



\bibitem{Ogawa07}
\bibinfo{author}{\bibfnamefont{N.}~\bibnamefont{Ogawa}},
  \bibinfo{author}{\bibfnamefont{G.}~\bibnamefont{Mikaelian}},
  \bibnamefont{and} \bibinfo{author}{\bibfnamefont{W.}~\bibnamefont{Ho}},
  \bibinfo{journal}{Phys. Rev. Lett.} \textbf{\bibinfo{volume}{98}},
  \bibinfo{pages}{166103} (\bibinfo{year}{2007}).

\bibitem{Schulze08}
\bibinfo{author}{\bibfnamefont{G.}~\bibnamefont{Schulze}},
  \bibinfo{author}{\bibfnamefont{K.~J.} \bibnamefont{Franke}},
  \bibinfo{author}{\bibfnamefont{A.}~\bibnamefont{Gagliardi}},
  \bibinfo{author}{\bibfnamefont{G.}~\bibnamefont{Romano}},
  \bibinfo{author}{\bibfnamefont{C.~S.} \bibnamefont{Lin}},
  \bibinfo{author}{\bibfnamefont{A.}~\bibnamefont{{Da Rosa}}},
  \bibinfo{author}{\bibfnamefont{T.~A.} \bibnamefont{Niehaus}},
  \bibinfo{author}{\bibfnamefont{T.}~\bibnamefont{Frauenheim}},
  \bibinfo{author}{\bibfnamefont{A.}~\bibnamefont{{Di Carlo}}},
  \bibinfo{author}{\bibfnamefont{A.}~\bibnamefont{Pecchia}}, \bibnamefont{and}
  \bibinfo{author}{\bibfnamefont{J.}~\bibnamefont{Pascual}},
  \bibinfo{journal}{Phys. Rev. Lett.} \textbf{\bibinfo{volume}{100}},
  \bibinfo{pages}{136801} (\bibinfo{year}{2008}).

\bibitem{Pump08}
\bibinfo{author}{\bibfnamefont{F.}~\bibnamefont{Pump}},
  \bibinfo{author}{\bibfnamefont{R.}~\bibnamefont{Temirov}},
  \bibinfo{author}{\bibfnamefont{O.}~\bibnamefont{Neucheva}},
  \bibinfo{author}{\bibfnamefont{S.}~\bibnamefont{Soubatch}},
  \bibinfo{author}{\bibfnamefont{S.}~\bibnamefont{Tautz}},
  \bibinfo{author}{\bibfnamefont{M.}~\bibnamefont{Rohlfing}}, \bibnamefont{and}
  \bibinfo{author}{\bibfnamefont{G.}~\bibnamefont{Cuniberti}},
  \bibinfo{journal}{Appl. Phys. A} \textbf{\bibinfo{volume}{93}},
  \bibinfo{pages}{335} (\bibinfo{year}{2008}).

\bibitem{Leon08}
\bibinfo{author}{\bibfnamefont{N.~P.} \bibnamefont{{de Leon}}},
  \bibinfo{author}{\bibfnamefont{W.}~\bibnamefont{Liang}},
  \bibinfo{author}{\bibfnamefont{Q.}~\bibnamefont{Gu}}, \bibnamefont{and}
  \bibinfo{author}{\bibfnamefont{H.}~\bibnamefont{Park}},
  \bibinfo{journal}{Nano Lett.} \textbf{\bibinfo{volume}{8}},
  \bibinfo{pages}{2963} (\bibinfo{year}{2008}).

\bibitem{Osorio10}
\bibinfo{author}{\bibfnamefont{E.~A.} \bibnamefont{Osorio}},
  \bibinfo{author}{\bibfnamefont{M.}~\bibnamefont{Ruben}},
  \bibinfo{author}{\bibfnamefont{J.~S.} \bibnamefont{Seldenthuis}},
  \bibinfo{author}{\bibfnamefont{J.~M.} \bibnamefont{Lehn}}, \bibnamefont{and}
  \bibinfo{author}{\bibfnamefont{H.~S.~J.} \bibnamefont{van~der Zant}},
  \bibinfo{journal}{Small} \textbf{\bibinfo{volume}{6}}, \bibinfo{pages}{174}
  (\bibinfo{year}{2010}).

\bibitem{Martin10}
\bibinfo{author}{\bibfnamefont{C.~A.} \bibnamefont{Martin}},
  \bibinfo{author}{\bibfnamefont{J.~M.} \bibnamefont{van Ruitenbeek}},
  \bibnamefont{and} \bibinfo{author}{\bibfnamefont{H.~S.~J.}
  \bibnamefont{van~de Zant}}, \bibinfo{journal}{Nanotechnology}
  \textbf{\bibinfo{volume}{21}}, \bibinfo{pages}{265201}
  (\bibinfo{year}{2010}).

\bibitem{lia02:725}
\bibinfo{author}{\bibfnamefont{W.}~\bibnamefont{Liang}},
  \bibinfo{author}{\bibfnamefont{M.}~\bibnamefont{Shores}},
  \bibinfo{author}{\bibfnamefont{M.}~\bibnamefont{Bockrath}},
  \bibinfo{author}{\bibfnamefont{J.}~\bibnamefont{Long}}, \bibnamefont{and}
  \bibinfo{author}{\bibfnamefont{H.}~\bibnamefont{Park}},
  \bibinfo{journal}{Nature (London)} \textbf{\bibinfo{volume}{417}},
  \bibinfo{pages}{725} (\bibinfo{year}{2002}).

\bibitem{che99:1550}
\bibinfo{author}{\bibfnamefont{J.}~\bibnamefont{Chen}},
  \bibinfo{author}{\bibfnamefont{M.}~\bibnamefont{Reed}},
  \bibinfo{author}{\bibfnamefont{A.}~\bibnamefont{Rawlett}}, \bibnamefont{and}
  \bibinfo{author}{\bibfnamefont{J.}~\bibnamefont{Tour}},
  \bibinfo{journal}{Science} \textbf{\bibinfo{volume}{286}},
  \bibinfo{pages}{1550} (\bibinfo{year}{1999}).

\bibitem{Gaudioso00}
\bibinfo{author}{\bibfnamefont{J.}~\bibnamefont{Gaudioso}},
  \bibinfo{author}{\bibfnamefont{L.~J.} \bibnamefont{Lauhon}},
  \bibnamefont{and} \bibinfo{author}{\bibfnamefont{W.}~\bibnamefont{Ho}},
  \bibinfo{journal}{Phys. Rev. Lett.} \textbf{\bibinfo{volume}{85}},
  \bibinfo{pages}{1918} (\bibinfo{year}{2000}).

\bibitem{blu05:167}
\bibinfo{author}{\bibfnamefont{A.}~\bibnamefont{Blum}},
  \bibinfo{author}{\bibfnamefont{J.}~\bibnamefont{Kushmerick}},
  \bibinfo{author}{\bibfnamefont{D.}~\bibnamefont{Long}},
  \bibinfo{author}{\bibfnamefont{C.}~\bibnamefont{Patterson}},
  \bibinfo{author}{\bibfnamefont{J.}~\bibnamefont{Jang}},
  \bibinfo{author}{\bibfnamefont{J.}~\bibnamefont{Henderson}},
  \bibinfo{author}{\bibfnamefont{Y.}~\bibnamefont{Yao}},
  \bibinfo{author}{\bibfnamefont{J.}~\bibnamefont{Tour}},
  \bibinfo{author}{\bibfnamefont{R.}~\bibnamefont{Shashidhar}},
  \bibnamefont{and} \bibinfo{author}{\bibfnamefont{B.}~\bibnamefont{Ratna}},
  \bibinfo{journal}{Nat. Mater.} \textbf{\bibinfo{volume}{4}},
  \bibinfo{pages}{167} (\bibinfo{year}{2005}).

\bibitem{Riel06}
\bibinfo{author}{\bibfnamefont{E.}~\bibnamefont{L\"ortscher}},
  \bibinfo{author}{\bibfnamefont{J.~W.} \bibnamefont{Ciszek}},
  \bibinfo{author}{\bibfnamefont{J.}~\bibnamefont{Tour}}, \bibnamefont{and}
  \bibinfo{author}{\bibfnamefont{H.}~\bibnamefont{Riel}},
  \bibinfo{journal}{Small} \textbf{\bibinfo{volume}{2}}, \bibinfo{pages}{973}
  (\bibinfo{year}{2006}).

\bibitem{Choi06}
\bibinfo{author}{\bibfnamefont{B.-Y.} \bibnamefont{Choi}},
  \bibinfo{author}{\bibfnamefont{S.-J.} \bibnamefont{Kahng}},
  \bibinfo{author}{\bibfnamefont{S.}~\bibnamefont{Kim}},
  \bibinfo{author}{\bibfnamefont{H.}~\bibnamefont{Kim}},
  \bibinfo{author}{\bibfnamefont{H.}~\bibnamefont{Kim}},
  \bibinfo{author}{\bibfnamefont{Y.}~\bibnamefont{Song}},
  \bibinfo{author}{\bibfnamefont{J.}~\bibnamefont{Ihm}}, \bibnamefont{and}
  \bibinfo{author}{\bibfnamefont{Y.}~\bibnamefont{Kuk}},
  \bibinfo{journal}{Phys. Rev. Lett.} \textbf{\bibinfo{volume}{96}},
  \bibinfo{pages}{156106} (\bibinfo{year}{2006}).

\bibitem{Ballmann12}
\bibinfo{author}{\bibfnamefont{S.}~\bibnamefont{Ballmann}},
  \bibinfo{author}{\bibfnamefont{R.}~\bibnamefont{{H\"artle}}},
  \bibinfo{author}{\bibfnamefont{P.}~\bibnamefont{Coto}},
  \bibinfo{author}{\bibfnamefont{M.}~\bibnamefont{Elbing}},
  \bibinfo{author}{\bibfnamefont{M.}~\bibnamefont{Mayor}},
  \bibinfo{author}{\bibfnamefont{M.}~\bibnamefont{Bryce}},
  \bibinfo{author}{\bibfnamefont{M.}~\bibnamefont{Thoss}}, \bibnamefont{and}
  \bibinfo{author}{\bibfnamefont{H.~B.} \bibnamefont{Weber}},
  \bibinfo{journal}{Phys. Rev. Lett.} \textbf{\bibinfo{volume}{109}},
  \bibinfo{pages}{056801} (\bibinfo{year}{2012}).



\bibitem{Guedon_Molen_12}
\bibinfo{author}{\bibfnamefont{C.}~\bibnamefont{Guedon}},
  \bibinfo{author}{\bibfnamefont{H.}~\bibnamefont{Valkenier}},
  \bibinfo{author}{\bibfnamefont{T.}~\bibnamefont{Markussen}},
  \bibinfo{author}{\bibfnamefont{K.}~\bibnamefont{Thygesen}},
  \bibinfo{author}{\bibfnamefont{J.}~\bibnamefont{Hummelen}}, \bibnamefont{and}
  \bibinfo{author}{\bibfnamefont{S.}~\bibnamefont{van~der Molen}},
  \bibinfo{journal}{Nat. Nanotechnol.} \textbf{\bibinfo{volume}{7}},
  \bibinfo{pages}{305} (\bibinfo{year}{2012}).

\bibitem{Vazquez_Venkatamaran12}
\bibinfo{author}{\bibfnamefont{H.}~\bibnamefont{Vazquez}},
  \bibinfo{author}{\bibfnamefont{R.}~\bibnamefont{Skouta}},
  \bibinfo{author}{\bibfnamefont{S.}~\bibnamefont{Schneebeli}},
  \bibinfo{author}{\bibfnamefont{M.}~\bibnamefont{Kamenetska}},
  \bibinfo{author}{\bibfnamefont{R.}~\bibnamefont{Breslow}},
  \bibinfo{author}{\bibfnamefont{L.}~\bibnamefont{Venkataraman}},
  \bibnamefont{and}
  \bibinfo{author}{\bibfnamefont{M.}~\bibnamefont{Hybertsen}},
  \bibinfo{journal}{Nat. Nanotechnol.} \textbf{\bibinfo{volume}{7}},
  \bibinfo{pages}{663} (\bibinfo{year}{2012}).


\bibitem{Landauer57}
\bibinfo{author}{\bibfnamefont{R.}~\bibnamefont{Landauer}},
  \bibinfo{journal}{IBM J. Res. Dev.} \textbf{\bibinfo{volume}{1}},
  \bibinfo{pages}{223} (\bibinfo{year}{1951}).

\bibitem{Bonca95}
\bibinfo{author}{\bibfnamefont{J.}~\bibnamefont{Bonca}} \bibnamefont{and}
  \bibinfo{author}{\bibfnamefont{S.}~\bibnamefont{Trugmann}},
  \bibinfo{journal}{Phys. Rev. Lett.} \textbf{\bibinfo{volume}{75}},
  \bibinfo{pages}{2566} (\bibinfo{year}{1995}).

\bibitem{Ness01}
\bibinfo{author}{\bibfnamefont{H.}~\bibnamefont{Ness}},
  \bibinfo{author}{\bibfnamefont{S.}~\bibnamefont{Shevlin}}, \bibnamefont{and}
  \bibinfo{author}{\bibfnamefont{A.}~\bibnamefont{Fisher}},
  \bibinfo{journal}{Phys. Rev. B} \textbf{\bibinfo{volume}{63}},
  \bibinfo{pages}{125422} (\bibinfo{year}{2001}).

\bibitem{Cizek04}
\bibinfo{author}{\bibfnamefont{M.}~\bibnamefont{Cizek}},
  \bibinfo{author}{\bibfnamefont{M.}~\bibnamefont{Thoss}}, \bibnamefont{and}
  \bibinfo{author}{\bibfnamefont{W.}~\bibnamefont{Domcke}},
  \bibinfo{journal}{Phys. Rev. B} \textbf{\bibinfo{volume}{70}},
  \bibinfo{pages}{125406} (\bibinfo{year}{2004}).

\bibitem{Cizek05}
\bibinfo{author}{\bibfnamefont{M.}~\bibnamefont{Cizek}},
  \bibinfo{author}{\bibfnamefont{M.}~\bibnamefont{Thoss}}, \bibnamefont{and}
  \bibinfo{author}{\bibfnamefont{W.}~\bibnamefont{Domcke}},
  \bibinfo{journal}{Czech.\ J.\ Phys.} \textbf{\bibinfo{volume}{55}},
  \bibinfo{pages}{189} (\bibinfo{year}{2005}).

\bibitem{Toroker07}
\bibinfo{author}{\bibfnamefont{M.}~\bibnamefont{Caspary-Toroker}}
  \bibnamefont{and} \bibinfo{author}{\bibfnamefont{U.}~\bibnamefont{Peskin}},
  \bibinfo{journal}{J. Chem. Phys.} \textbf{\bibinfo{volume}{127}},
  \bibinfo{pages}{154706} (\bibinfo{year}{2007}).

\bibitem{Benesch08}
\bibinfo{author}{\bibfnamefont{C.}~\bibnamefont{Benesch}},
  \bibinfo{author}{\bibfnamefont{M.}~\bibnamefont{Cizek}},
  \bibinfo{author}{\bibfnamefont{J.}~\bibnamefont{Klimes}},
  \bibinfo{author}{\bibfnamefont{I.}~\bibnamefont{Kondov}},
  \bibinfo{author}{\bibfnamefont{M.}~\bibnamefont{Thoss}}, \bibnamefont{and}
  \bibinfo{author}{\bibfnamefont{W.}~\bibnamefont{Domcke}},
  \bibinfo{journal}{J. Phys. Chem. C} \textbf{\bibinfo{volume}{112}},
  \bibinfo{pages}{9880} (\bibinfo{year}{2008}).

\bibitem{Zimbovskaya09}
\bibinfo{author}{\bibfnamefont{N.~A.} \bibnamefont{Zimbovskaya}}
  \bibnamefont{and} \bibinfo{author}{\bibfnamefont{M.~M.}
  \bibnamefont{Kuklja}}, \bibinfo{journal}{J. Chem. Phys.}
  \textbf{\bibinfo{volume}{131}}, \bibinfo{pages}{114703}
  (\bibinfo{year}{2009}).

\bibitem{Seidemann10}
\bibinfo{author}{\bibfnamefont{R.}~\bibnamefont{Jorn}} \bibnamefont{and}
  \bibinfo{author}{\bibfnamefont{T.}~\bibnamefont{Seidemann}},
  \bibinfo{journal}{J. Chem. Phys.} \textbf{\bibinfo{volume}{131}},
  \bibinfo{pages}{244114} (\bibinfo{year}{2009}).

\bibitem{Flensberg03}
\bibinfo{author}{\bibfnamefont{K.}~\bibnamefont{Flensberg}},
  \bibinfo{journal}{Phys. Rev. B} \textbf{\bibinfo{volume}{68}},
  \bibinfo{pages}{205323} (\bibinfo{year}{2003}).

\bibitem{Mitra04}
\bibinfo{author}{\bibfnamefont{A.}~\bibnamefont{Mitra}},
  \bibinfo{author}{\bibfnamefont{I.}~\bibnamefont{Aleiner}}, \bibnamefont{and}
  \bibinfo{author}{\bibfnamefont{A.~J.} \bibnamefont{Millis}},
  \bibinfo{journal}{Phys. Rev. B} \textbf{\bibinfo{volume}{69}},
  \bibinfo{pages}{245302} (\bibinfo{year}{2004}).

\bibitem{Galperin06}
\bibinfo{author}{\bibfnamefont{M.}~\bibnamefont{Galperin}},
  \bibinfo{author}{\bibfnamefont{M.}~\bibnamefont{Ratner}}, \bibnamefont{and}
  \bibinfo{author}{\bibfnamefont{A.}~\bibnamefont{Nitzan}},
  \bibinfo{journal}{Phys. Rev. B} \textbf{\bibinfo{volume}{73}},
  \bibinfo{pages}{045314} (\bibinfo{year}{2006}).

\bibitem{Ryndyk06}
\bibinfo{author}{\bibfnamefont{D.~A.} \bibnamefont{Ryndyk}},
  \bibinfo{author}{\bibfnamefont{M.}~\bibnamefont{Hartung}}, \bibnamefont{and}
  \bibinfo{author}{\bibfnamefont{G.}~\bibnamefont{Cuniberti}},
  \bibinfo{journal}{Phys. Rev. B} \textbf{\bibinfo{volume}{73}},
  \bibinfo{pages}{045420} (\bibinfo{year}{2006}).

\bibitem{Frederiksen07}
\bibinfo{author}{\bibfnamefont{T.}~\bibnamefont{Frederiksen}},
  \bibinfo{author}{\bibfnamefont{M.}~\bibnamefont{Paulsson}},
  \bibinfo{author}{\bibfnamefont{M.}~\bibnamefont{Brandbyge}},
  \bibnamefont{and} \bibinfo{author}{\bibfnamefont{A.}~\bibnamefont{Jauho}},
  \bibinfo{journal}{Phys. Rev. B} \textbf{\bibinfo{volume}{75}},
  \bibinfo{pages}{205413} (\bibinfo{year}{2007}).

\bibitem{Tahir08}
\bibinfo{author}{\bibfnamefont{M.}~\bibnamefont{Tahir}} \bibnamefont{and}
  \bibinfo{author}{\bibfnamefont{A.}~\bibnamefont{MacKinnon}},
  \bibinfo{journal}{Phys. Rev. B} \textbf{\bibinfo{volume}{77}},
  \bibinfo{pages}{224305} (\bibinfo{year}{2008}).

\bibitem{Haertle08}
\bibinfo{author}{\bibfnamefont{R.}~\bibnamefont{H{\"a}rtle}},
  \bibinfo{author}{\bibfnamefont{C.}~\bibnamefont{Benesch}}, \bibnamefont{and}
  \bibinfo{author}{\bibfnamefont{M.}~\bibnamefont{Thoss}},
  \bibinfo{journal}{Phys. Rev. B} \textbf{\bibinfo{volume}{77}},
  \bibinfo{pages}{205314} (\bibinfo{year}{2008}).

\bibitem{Stafford09}
\bibinfo{author}{\bibfnamefont{J.~P.} \bibnamefont{Bergfield}}
  \bibnamefont{and} \bibinfo{author}{\bibfnamefont{C.~A.}
  \bibnamefont{Stafford}}, \bibinfo{journal}{Phys. Rev. B}
  \textbf{\bibinfo{volume}{79}}, \bibinfo{pages}{245125}
  (\bibinfo{year}{2009}).

\bibitem{Haertle09}
\bibinfo{author}{\bibfnamefont{R.}~\bibnamefont{H{\"a}rtle}},
  \bibinfo{author}{\bibfnamefont{C.}~\bibnamefont{Benesch}}, \bibnamefont{and}
  \bibinfo{author}{\bibfnamefont{M.}~\bibnamefont{Thoss}},
  \bibinfo{journal}{Phys. Rev. Lett.} \textbf{\bibinfo{volume}{102}},
  \bibinfo{pages}{146801} (\bibinfo{year}{2009}).

\bibitem{May02}
\bibinfo{author}{\bibfnamefont{V.}~\bibnamefont{May}}, \bibinfo{journal}{Phys.
  Rev. B} \textbf{\bibinfo{volume}{66}}, \bibinfo{pages}{245411}
  (\bibinfo{year}{2002}).

\bibitem{Lehmann04}
\bibinfo{author}{\bibfnamefont{J.}~\bibnamefont{Lehmann}},
  \bibinfo{author}{\bibfnamefont{S.}~\bibnamefont{Kohler}},
  \bibinfo{author}{\bibfnamefont{V.}~\bibnamefont{May}}, \bibnamefont{and}
  \bibinfo{author}{\bibfnamefont{P.}~\bibnamefont{{H\"anggi}}},
  \bibinfo{journal}{J. Chem. Phys.} \textbf{\bibinfo{volume}{121}},
  \bibinfo{pages}{2278} (\bibinfo{year}{2004}).

\bibitem{Pedersen05}
\bibinfo{author}{\bibfnamefont{J.~N.} \bibnamefont{Pedersen}} \bibnamefont{and}
  \bibinfo{author}{\bibfnamefont{A.}~\bibnamefont{Wacker}},
  \bibinfo{journal}{Phys. Rev. B} \textbf{\bibinfo{volume}{72}},
  \bibinfo{pages}{195330} (\bibinfo{year}{2005}).

\bibitem{Harbola06}
\bibinfo{author}{\bibfnamefont{U.}~\bibnamefont{Harbola}},
  \bibinfo{author}{\bibfnamefont{M.}~\bibnamefont{Esposito}}, \bibnamefont{and}
  \bibinfo{author}{\bibfnamefont{S.}~\bibnamefont{Mukamel}},
  \bibinfo{journal}{Phys. Rev. B} \textbf{\bibinfo{volume}{74}},
  \bibinfo{pages}{235309} (\bibinfo{year}{2006}).

\bibitem{Zazunov06}
\bibinfo{author}{\bibfnamefont{A.}~\bibnamefont{Zazunov}},
  \bibinfo{author}{\bibfnamefont{D.}~\bibnamefont{Feinberg}}, \bibnamefont{and}
  \bibinfo{author}{\bibfnamefont{T.}~\bibnamefont{Martin}},
  \bibinfo{journal}{Phys. Rev. B} \textbf{\bibinfo{volume}{73}},
  \bibinfo{pages}{115405} (\bibinfo{year}{2006}).

\bibitem{Siddiqui07}
\bibinfo{author}{\bibfnamefont{L.}~\bibnamefont{Siddiqui}},
  \bibinfo{author}{\bibfnamefont{A.~W.} \bibnamefont{Ghosh}}, \bibnamefont{and}
  \bibinfo{author}{\bibfnamefont{S.}~\bibnamefont{Datta}},
  \bibinfo{journal}{Phys. Rev. B} \textbf{\bibinfo{volume}{76}},
  \bibinfo{pages}{085433} (\bibinfo{year}{2007}).

\bibitem{Timm08}
\bibinfo{author}{\bibfnamefont{C.}~\bibnamefont{Timm}}, \bibinfo{journal}{Phys.
  Rev. B} \textbf{\bibinfo{volume}{77}}, \bibinfo{pages}{195416}
  (\bibinfo{year}{2008}).

\bibitem{May08}
\bibinfo{author}{\bibfnamefont{V.}~\bibnamefont{May}} \bibnamefont{and}
  \bibinfo{author}{\bibfnamefont{O.}~\bibnamefont{K\"uhn}},
  \bibinfo{journal}{Phys. Rev. B} \textbf{\bibinfo{volume}{77}},
  \bibinfo{pages}{115439} (\bibinfo{year}{2008}).

\bibitem{May08b}
\bibinfo{author}{\bibfnamefont{V.}~\bibnamefont{May}} \bibnamefont{and}
  \bibinfo{author}{\bibfnamefont{O.}~\bibnamefont{K\"uhn}},
  \bibinfo{journal}{Phys. Rev. B} \textbf{\bibinfo{volume}{77}},
  \bibinfo{pages}{115440} (\bibinfo{year}{2008}).

\bibitem{Leijnse09}
\bibinfo{author}{\bibfnamefont{M.}~\bibnamefont{Leijnse}} \bibnamefont{and}
  \bibinfo{author}{\bibfnamefont{M.~R.} \bibnamefont{Wegewijs}},
  \bibinfo{journal}{Phys. Rev. B} \textbf{\bibinfo{volume}{78}},
  \bibinfo{pages}{235424} (\bibinfo{year}{2008}).

\bibitem{Esposito09}
\bibinfo{author}{\bibfnamefont{M.}~\bibnamefont{Esposito}} \bibnamefont{and}
  \bibinfo{author}{\bibfnamefont{M.}~\bibnamefont{Galperin}},
  \bibinfo{journal}{Phys. Rev. B} \textbf{\bibinfo{volume}{79}},
  \bibinfo{pages}{205303} (\bibinfo{year}{2009}).

\bibitem{Haertle11}
\bibinfo{author}{\bibfnamefont{R.}~\bibnamefont{{H\"artle}}} \bibnamefont{and}
  \bibinfo{author}{\bibfnamefont{M.}~\bibnamefont{Thoss}},
  \bibinfo{journal}{Phys. Rev. B} \textbf{\bibinfo{volume}{83}},
  \bibinfo{pages}{115414} (\bibinfo{year}{2011}).

\bibitem{muh08:176403}
\bibinfo{author}{\bibfnamefont{L.}~\bibnamefont{{M\"uhlbacher}}}
  \bibnamefont{and} \bibinfo{author}{\bibfnamefont{E.}~\bibnamefont{Rabani}},
  \bibinfo{journal}{Phys. Rev. Lett.} \textbf{\bibinfo{volume}{100}},
  \bibinfo{pages}{176403} (\bibinfo{year}{2008}).

\bibitem{wei08:195316}
\bibinfo{author}{\bibfnamefont{S.}~\bibnamefont{Weiss}},
  \bibinfo{author}{\bibfnamefont{J.}~\bibnamefont{Eckel}},
  \bibinfo{author}{\bibfnamefont{M.}~\bibnamefont{Thorwart}}, \bibnamefont{and}
  \bibinfo{author}{\bibfnamefont{R.}~\bibnamefont{Egger}},
  \bibinfo{journal}{Phys. Rev. B} \textbf{\bibinfo{volume}{77}},
  \bibinfo{pages}{195316} (\bibinfo{year}{2008}).

\bibitem{Segal10}
\bibinfo{author}{\bibfnamefont{D.}~\bibnamefont{Segal}},
  \bibinfo{author}{\bibnamefont{A.J.Millis}}, \bibnamefont{and}
  \bibinfo{author}{\bibfnamefont{D.}~\bibnamefont{Reichman}},
  \bibinfo{journal}{Phys. Rev. B} \textbf{\bibinfo{volume}{82}},
  \bibinfo{pages}{205323} (\bibinfo{year}{2010}).

\bibitem{Werner09}
\bibinfo{author}{\bibfnamefont{P.}~\bibnamefont{Werner}},
  \bibinfo{author}{\bibfnamefont{T.}~\bibnamefont{Oka}}, \bibnamefont{and}
  \bibinfo{author}{\bibfnamefont{A.~J.} \bibnamefont{Millis}},
  \bibinfo{journal}{Phys. Rev. B} \textbf{\bibinfo{volume}{79}},
  \bibinfo{pages}{035320} (\bibinfo{year}{2009}).

\bibitem{Schiro09}
\bibinfo{author}{\bibfnamefont{M.}~\bibnamefont{Schiro}} \bibnamefont{and}
  \bibinfo{author}{\bibfnamefont{M.}~\bibnamefont{Fabrizio}},
  \bibinfo{journal}{Phys. Rev. B} \textbf{\bibinfo{volume}{79}},
  \bibinfo{pages}{153302} (\bibinfo{year}{2009}).

\bibitem{and08:066804}
\bibinfo{author}{\bibfnamefont{F.~B.} \bibnamefont{Anders}},
  \bibinfo{journal}{Phys. Rev. Lett.} \textbf{\bibinfo{volume}{101}},
  \bibinfo{pages}{066804} (\bibinfo{year}{2008}).

\bibitem{wan09:024114}
\bibinfo{author}{\bibfnamefont{H.}~\bibnamefont{Wang}} \bibnamefont{and}
  \bibinfo{author}{\bibfnamefont{M.}~\bibnamefont{Thoss}}, \bibinfo{journal}{J.
  Chem. Phys.} \textbf{\bibinfo{volume}{131}}, \bibinfo{pages}{024114}
  (\bibinfo{year}{2009}).

\bibitem{wan11:244506}
\bibinfo{author}{\bibfnamefont{H.}~\bibnamefont{Wang}},
  \bibinfo{author}{\bibfnamefont{I.}~\bibnamefont{Pshenichnyuk}},
  \bibinfo{author}{\bibfnamefont{R.}~\bibnamefont{H\"{a}rtle}},
  \bibnamefont{and} \bibinfo{author}{\bibfnamefont{M.}~\bibnamefont{Thoss}},
  \bibinfo{journal}{J. Chem. Phys.} \textbf{\bibinfo{volume}{135}},
  \bibinfo{pages}{244506} (\bibinfo{year}{2011}).

\bibitem{wan13:134704}
\bibinfo{author}{\bibfnamefont{H.}~\bibnamefont{Wang}} \bibnamefont{and}
  \bibinfo{author}{\bibfnamefont{M.}~\bibnamefont{Thoss}}, \bibinfo{journal}{J.
  Chem. Phys.} \textbf{\bibinfo{volume}{138}}, \bibinfo{pages}{134704}
  (\bibinfo{year}{2013}).

\bibitem{wan13:7431}
\bibinfo{author}{\bibfnamefont{H.}~\bibnamefont{Wang}} \bibnamefont{and}
  \bibinfo{author}{\bibfnamefont{M.}~\bibnamefont{Thoss}}, \bibinfo{journal}{J.
  Phys. Chem. A} \textbf{\bibinfo{volume}{117}}, \bibinfo{pages}{7431}
  (\bibinfo{year}{2013}).

\bibitem{wil13:045137}
\bibinfo{author}{\bibfnamefont{E.~Y.} \bibnamefont{Wilner}},
  \bibinfo{author}{\bibfnamefont{H.}~\bibnamefont{Wang}},
  \bibinfo{author}{\bibfnamefont{G.}~\bibnamefont{Cohen}},
  \bibinfo{author}{\bibfnamefont{M.}~\bibnamefont{Thoss}}, \bibnamefont{and}
  \bibinfo{author}{\bibfnamefont{E.}~\bibnamefont{Rabani}},
  \bibinfo{journal}{Phys. Rev. B} \textbf{\bibinfo{volume}{88}},
  \bibinfo{pages}{045137} (\bibinfo{year}{2013}).

\bibitem{sti98:1732}
\bibinfo{author}{\bibfnamefont{B.}~\bibnamefont{Stipe}},
  \bibinfo{author}{\bibfnamefont{M.}~\bibnamefont{Rezai}}, \bibnamefont{and}
  \bibinfo{author}{\bibfnamefont{W.}~\bibnamefont{Ho}},
  \bibinfo{journal}{Science} \textbf{\bibinfo{volume}{280}},
  \bibinfo{pages}{1732} (\bibinfo{year}{1998}).

\bibitem{Natelson04}
\bibinfo{author}{\bibfnamefont{L.~H.} \bibnamefont{Yu}},
  \bibinfo{author}{\bibfnamefont{Z.~K.} \bibnamefont{Keane}},
  \bibinfo{author}{\bibfnamefont{J.~W.} \bibnamefont{Ciszek}},
  \bibinfo{author}{\bibfnamefont{L.}~\bibnamefont{Cheng}},
  \bibinfo{author}{\bibfnamefont{M.~P.} \bibnamefont{Stewart}},
  \bibinfo{author}{\bibfnamefont{J.~M.} \bibnamefont{Tour}}, \bibnamefont{and}
  \bibinfo{author}{\bibfnamefont{D.}~\bibnamefont{Natelson}},
  \bibinfo{journal}{Phys. Rev. Lett.} \textbf{\bibinfo{volume}{93}},
  \bibinfo{pages}{266802} (\bibinfo{year}{2004}).

\bibitem{kus04:639}
\bibinfo{author}{\bibfnamefont{J.}~\bibnamefont{Kushmerick}},
  \bibinfo{author}{\bibfnamefont{J.}~\bibnamefont{Lazorcik}},
  \bibinfo{author}{\bibfnamefont{C.}~\bibnamefont{Patterson}},
  \bibinfo{author}{\bibfnamefont{R.}~\bibnamefont{Shashidhar}},
  \bibinfo{author}{\bibfnamefont{D.~S.} \bibnamefont{Seferos}},
  \bibnamefont{and} \bibinfo{author}{\bibfnamefont{G.~C.} \bibnamefont{Bazan}},
  \bibinfo{journal}{Nano Lett.} \textbf{\bibinfo{volume}{4}},
  \bibinfo{pages}{639} (\bibinfo{year}{2004}).

\bibitem{pas05:203}
\bibinfo{author}{\bibfnamefont{A.}~\bibnamefont{Pasupathy}},
  \bibinfo{author}{\bibfnamefont{J.}~\bibnamefont{Park}},
  \bibinfo{author}{\bibfnamefont{C.}~\bibnamefont{Chang}},
  \bibinfo{author}{\bibfnamefont{A.}~\bibnamefont{Soldatov}},
  \bibinfo{author}{\bibfnamefont{S.}~\bibnamefont{Lebedkin}},
  \bibinfo{author}{\bibfnamefont{R.}~\bibnamefont{Bialczak}},
  \bibinfo{author}{\bibfnamefont{J.}~\bibnamefont{Grose}},
  \bibinfo{author}{\bibfnamefont{L.}~\bibnamefont{Donev}},
  \bibinfo{author}{\bibfnamefont{J.}~\bibnamefont{Sethna}},
  \bibinfo{author}{\bibfnamefont{D.}~\bibnamefont{Ralph}}, \bibnamefont{and}
  \bibinfo{author}{\bibfnamefont{P.}~\bibnamefont{McEuen}},
  \bibinfo{journal}{Nano Lett.} \textbf{\bibinfo{volume}{5}},
  \bibinfo{pages}{203} (\bibinfo{year}{2005}).

\bibitem{Sapmaz06}
\bibinfo{author}{\bibfnamefont{S.}~\bibnamefont{Sapmaz}},
  \bibinfo{author}{\bibfnamefont{P.}~\bibnamefont{Jarillo-Herrero}},
  \bibinfo{author}{\bibfnamefont{Y.~M.} \bibnamefont{Blanter}},
  \bibinfo{author}{\bibfnamefont{C.}~\bibnamefont{Dekker}}, \bibnamefont{and}
  \bibinfo{author}{\bibfnamefont{H.~S.} \bibnamefont{{van der Zant}}},
  \bibinfo{journal}{Phys. Rev. Lett.} \textbf{\bibinfo{volume}{96}},
  \bibinfo{pages}{026801} (\bibinfo{year}{2006}).

\bibitem{Thijssen06}
\bibinfo{author}{\bibfnamefont{W.~H.~A.} \bibnamefont{Thijssen}},
  \bibinfo{author}{\bibfnamefont{D.}~\bibnamefont{Djukic}},
  \bibinfo{author}{\bibfnamefont{A.~F.} \bibnamefont{Otte}},
  \bibinfo{author}{\bibfnamefont{R.~H.} \bibnamefont{Bremmer}},
  \bibnamefont{and} \bibinfo{author}{\bibfnamefont{J.~M.} \bibnamefont{van
  Ruitenbeek}}, \bibinfo{journal}{Phys. Rev. Lett.}
  \textbf{\bibinfo{volume}{97}}, \bibinfo{pages}{226806}
  (\bibinfo{year}{2006}).

\bibitem{Parks07}
\bibinfo{author}{\bibfnamefont{J.}~\bibnamefont{Parks}},
  \bibinfo{author}{\bibfnamefont{A.}~\bibnamefont{Champagne}},
  \bibinfo{author}{\bibfnamefont{G.}~\bibnamefont{Hutchison}},
  \bibinfo{author}{\bibfnamefont{S.}~\bibnamefont{Flores-Torres}},
  \bibinfo{author}{\bibfnamefont{H.}~\bibnamefont{Abruna}}, \bibnamefont{and}
  \bibinfo{author}{\bibfnamefont{D.}~\bibnamefont{Ralph}},
  \bibinfo{journal}{Phys. Rev. Lett.} \textbf{\bibinfo{volume}{99}},
  \bibinfo{pages}{026601} (\bibinfo{year}{2007}).

\bibitem{Boehler07}
\bibinfo{author}{\bibfnamefont{T.}~\bibnamefont{{B\"ohler}}},
  \bibinfo{author}{\bibfnamefont{A.}~\bibnamefont{Edtbauer}}, \bibnamefont{and}
  \bibinfo{author}{\bibfnamefont{E.}~\bibnamefont{Scheer}},
  \bibinfo{journal}{Phys. Rev. B} \textbf{\bibinfo{volume}{76}},
  \bibinfo{pages}{125432} (\bibinfo{year}{2007}).

\bibitem{Huang06}
\bibinfo{author}{\bibfnamefont{Z.}~\bibnamefont{Huang}},
  \bibinfo{author}{\bibfnamefont{B.}~\bibnamefont{Xu}},
  \bibinfo{author}{\bibfnamefont{Y.}~\bibnamefont{Chen}},
  \bibinfo{author}{\bibfnamefont{M.~D.} \bibnamefont{Ventra}},
  \bibnamefont{and} \bibinfo{author}{\bibfnamefont{N.}~\bibnamefont{Tao}},
  \bibinfo{journal}{Nano Lett.} \textbf{\bibinfo{volume}{6}},
  \bibinfo{pages}{1240} (\bibinfo{year}{2006}).

\bibitem{Natelson08}
\bibinfo{author}{\bibfnamefont{D.~R.} \bibnamefont{Ward}},
  \bibinfo{author}{\bibfnamefont{N.~J.} \bibnamefont{Halas}},
  \bibinfo{author}{\bibfnamefont{J.~W.} \bibnamefont{Ciszek}},
  \bibinfo{author}{\bibfnamefont{J.~M.} \bibnamefont{Tour}},
  \bibinfo{author}{\bibfnamefont{Y.}~\bibnamefont{Wu}},
  \bibinfo{author}{\bibfnamefont{P.}~\bibnamefont{Nordlander}},
  \bibnamefont{and} \bibinfo{author}{\bibfnamefont{D.}~\bibnamefont{Natelson}},
  \bibinfo{journal}{Nano Lett.} \textbf{\bibinfo{volume}{8}},
  \bibinfo{pages}{919} (\bibinfo{year}{2008}).

\bibitem{Ioffe08}
\bibinfo{author}{\bibfnamefont{Z.}~\bibnamefont{Ioffe}},
  \bibinfo{author}{\bibfnamefont{T.}~\bibnamefont{Shamai}},
  \bibinfo{author}{\bibfnamefont{A.}~\bibnamefont{Ophir}},
  \bibinfo{author}{\bibfnamefont{G.}~\bibnamefont{Noy}},
  \bibinfo{author}{\bibfnamefont{I.}~\bibnamefont{Yutsis}},
  \bibinfo{author}{\bibfnamefont{K.}~\bibnamefont{Kfir}},
  \bibinfo{author}{\bibfnamefont{O.}~\bibnamefont{Cheshnovsky}},
  \bibnamefont{and} \bibinfo{author}{\bibfnamefont{Y.}~\bibnamefont{Selzer}},
  \bibinfo{journal}{Nature Nanotech.} \textbf{\bibinfo{volume}{3}},
  \bibinfo{pages}{727} (\bibinfo{year}{2008}).

\bibitem{Huettel09}
\bibinfo{author}{\bibfnamefont{A.~K.} \bibnamefont{H\"uttel}},
  \bibinfo{author}{\bibfnamefont{B.}~\bibnamefont{Witkamp}},
  \bibinfo{author}{\bibfnamefont{M.}~\bibnamefont{Leijnse}},
  \bibinfo{author}{\bibfnamefont{M.~R.} \bibnamefont{Wegewijs}},
  \bibnamefont{and} \bibinfo{author}{\bibfnamefont{H.~S.~J.}
  \bibnamefont{van~der Zant}}, \bibinfo{journal}{Phys. Rev. Lett.}
  \textbf{\bibinfo{volume}{102}}, \bibinfo{pages}{225501}
  (\bibinfo{year}{2009}).

\bibitem{Ballmann10}
\bibinfo{author}{\bibfnamefont{S.}~\bibnamefont{Ballmann}},
  \bibinfo{author}{\bibfnamefont{W.}~\bibnamefont{Hieringer}},
  \bibinfo{author}{\bibfnamefont{D.}~\bibnamefont{Secker}},
  \bibinfo{author}{\bibfnamefont{Q.}~\bibnamefont{Zheng}},
  \bibinfo{author}{\bibfnamefont{J.~A.} \bibnamefont{Gladysz}},
  \bibinfo{author}{\bibfnamefont{A.}~\bibnamefont{G\"orling}},
  \bibnamefont{and} \bibinfo{author}{\bibfnamefont{H.~B.} \bibnamefont{Weber}},
  \bibinfo{journal}{Chem. Phys. Chem.} \textbf{\bibinfo{volume}{11}},
  \bibinfo{pages}{2256} (\bibinfo{year}{2010}).

\bibitem{Secker11}
\bibinfo{author}{\bibfnamefont{D.}~\bibnamefont{Secker}},
  \bibinfo{author}{\bibfnamefont{S.}~\bibnamefont{Wagner}},
  \bibinfo{author}{\bibfnamefont{S.}~\bibnamefont{Ballmann}},
  \bibinfo{author}{\bibfnamefont{R.}~\bibnamefont{{H\"artle}}},
  \bibinfo{author}{\bibfnamefont{M.}~\bibnamefont{Thoss}}, \bibnamefont{and}
  \bibinfo{author}{\bibfnamefont{H.~B.} \bibnamefont{Weber}},
  \bibinfo{journal}{Phys. Rev. Lett.} \textbf{\bibinfo{volume}{106}},
  \bibinfo{pages}{136807} (\bibinfo{year}{2011}).

\bibitem{Ballmann13}
\bibinfo{author}{\bibfnamefont{S.}~\bibnamefont{Ballmann}},
  \bibinfo{author}{\bibfnamefont{W.}~\bibnamefont{Hieringer}},
  \bibinfo{author}{\bibfnamefont{R.}~\bibnamefont{{H\"artle}}},
  \bibinfo{author}{\bibfnamefont{P.}~\bibnamefont{Coto}},
  \bibinfo{author}{\bibfnamefont{M.}~\bibnamefont{Bryce}},
  \bibinfo{author}{\bibfnamefont{A.}~\bibnamefont{G\"orling}},
  \bibinfo{author}{\bibfnamefont{M.}~\bibnamefont{Thoss}}, \bibnamefont{and}
  \bibinfo{author}{\bibfnamefont{H.~B.} \bibnamefont{Weber}},
  \bibinfo{journal}{Phys. Status Solidi B} \textbf{\bibinfo{volume}{250}},
  \bibinfo{pages}{2452} (\bibinfo{year}{2013}).

\bibitem{zha13:164121}
\bibinfo{author}{\bibfnamefont{Y.}~\bibnamefont{Zhang}},
  \bibinfo{author}{\bibfnamefont{C.~Y.} \bibnamefont{Yam}}, \bibnamefont{and}
  \bibinfo{author}{\bibfnamefont{G.}~\bibnamefont{Chen}}, \bibinfo{journal}{J.
  Chem. Phys.} \textbf{\bibinfo{volume}{138}}, \bibinfo{pages}{164121}
  (\bibinfo{year}{2013}).

\bibitem{Verdozzi06}
\bibinfo{author}{\bibfnamefont{C.}~\bibnamefont{Verdozzi}},
  \bibinfo{author}{\bibfnamefont{G.}~\bibnamefont{Stefanucci}},
  \bibnamefont{and} \bibinfo{author}{\bibfnamefont{C.}~\bibnamefont{Almbladh}},
  \bibinfo{journal}{Phys. Rev. Lett.} \textbf{\bibinfo{volume}{97}},
  \bibinfo{pages}{046603} (\bibinfo{year}{2006}).

\bibitem{leg87:1}
\bibinfo{author}{\bibfnamefont{A.~J.} \bibnamefont{Leggett}},
  \bibinfo{author}{\bibfnamefont{S.}~\bibnamefont{Chakravarty}},
  \bibinfo{author}{\bibfnamefont{A.~T.} \bibnamefont{Dorsey}},
  \bibinfo{author}{\bibfnamefont{M.~P.~A.} \bibnamefont{Fisher}},
  \bibinfo{author}{\bibfnamefont{A.}~\bibnamefont{Garg}}, \bibnamefont{and}
  \bibinfo{author}{\bibfnamefont{W.}~\bibnamefont{Zwerger}},
  \bibinfo{journal}{Rev. Mod. Phys.}
  \textbf{\bibinfo{volume}{59}}(\bibinfo{number}{1}), \bibinfo{pages}{1}
  (\bibinfo{year}{1987}).

\bibitem{Weiss93}
\bibinfo{author}{\bibfnamefont{U.}~\bibnamefont{Weiss}},
  \emph{\bibinfo{title}{Quantum Dissipative Systems}}
  (\bibinfo{publisher}{World Scientific}, \bibinfo{address}{Singapore},
  \bibinfo{year}{1993}).

\bibitem{wan03:1289}
\bibinfo{author}{\bibfnamefont{H.}~\bibnamefont{Wang}} \bibnamefont{and}
  \bibinfo{author}{\bibfnamefont{M.}~\bibnamefont{Thoss}}, \bibinfo{journal}{J.
  Chem. Phys.} \textbf{\bibinfo{volume}{119}},
  \bibinfo{pages}{1289} (\bibinfo{year}{2003}).

\bibitem{fer12:081412}
\bibinfo{author}{\bibfnamefont{K.~F.} \bibnamefont{Albrecht}},
  \bibinfo{author}{\bibfnamefont{H.}~\bibnamefont{Wang}},
  \bibinfo{author}{\bibfnamefont{L.}~\bibnamefont{M\"{u}hlbacher}},
  \bibinfo{author}{\bibfnamefont{M.}~\bibnamefont{Thoss}}, \bibnamefont{and}
  \bibinfo{author}{\bibfnamefont{A.}~\bibnamefont{Komnik}},
  \bibinfo{journal}{Phys. Rev. B} \textbf{\bibinfo{volume}{86}},
  \bibinfo{pages}{081412} (\bibinfo{year}{2012}).

\bibitem{wan15:7951}
\bibinfo{author}{\bibfnamefont{H.}~\bibnamefont{Wang}}, \bibinfo{journal}{J.
  Phys. Chem. A} \textbf{\bibinfo{volume}{119}}, \bibinfo{pages}{7951}
  (\bibinfo{year}{2015}).

\bibitem{Meyer90}
\bibinfo{author}{\bibfnamefont{H.-D.} \bibnamefont{Meyer}},
  \bibinfo{author}{\bibfnamefont{U.}~\bibnamefont{Manthe}}, \bibnamefont{and}
  \bibinfo{author}{\bibfnamefont{L.}~\bibnamefont{Cederbaum}},
  \bibinfo{journal}{Chem. Phys. Lett.} \textbf{\bibinfo{volume}{165}},
  \bibinfo{pages}{73} (\bibinfo{year}{1990}).

\bibitem{Meyer09}
\bibinfo{author}{\bibfnamefont{H.-D.} \bibnamefont{Meyer}},
  \bibinfo{author}{\bibfnamefont{F.}~\bibnamefont{Gatti}}, \bibnamefont{and}
  \bibinfo{author}{\bibfnamefont{G.}~\bibnamefont{Worth}},
  \emph{\bibinfo{title}{Multidimensional Quantum Dynamics: MCTDH Theory and
  Applications}} (\bibinfo{publisher}{Whiley-VCH}, \bibinfo{address}{Weinheim},
  \bibinfo{year}{2009}).



\bibitem{tho06:210}
\bibinfo{author}{\bibfnamefont{M.}~\bibnamefont{Thoss}} \bibnamefont{and}
  \bibinfo{author}{\bibfnamefont{H.}~\bibnamefont{Wang}},
  \bibinfo{journal}{Chem. Phys.}
  \textbf{\bibinfo{volume}{322}}, \bibinfo{pages}{210}
  (\bibinfo{year}{2006}).

\bibitem{wan06:034114}
\bibinfo{author}{\bibfnamefont{H.}~\bibnamefont{Wang}} \bibnamefont{and}
  \bibinfo{author}{\bibfnamefont{M.}~\bibnamefont{Thoss}}, \bibinfo{journal}{J.
  Chem. Phys.} \textbf{\bibinfo{volume}{124}},
  \bibinfo{pages}{034114} (\bibinfo{year}{2006}).

\bibitem{kon06:1364}
\bibinfo{author}{\bibfnamefont{I.}~\bibnamefont{Kondov}},
  \bibinfo{author}{\bibfnamefont{H.}~\bibnamefont{Wang}}, \bibnamefont{and}
  \bibinfo{author}{\bibfnamefont{M.}~\bibnamefont{Thoss}}, \bibinfo{journal}{J.
  Phys. Chem. A} \textbf{\bibinfo{volume}{110}},
  \bibinfo{pages}{1364} (\bibinfo{year}{2006}).

\bibitem{wan07:10369}
\bibinfo{author}{\bibfnamefont{H.}~\bibnamefont{Wang}} \bibnamefont{and}
  \bibinfo{author}{\bibfnamefont{M.}~\bibnamefont{Thoss}}, \bibinfo{journal}{J.
  Phys. Chem. A} \textbf{\bibinfo{volume}{111}}, \bibinfo{pages}{10369}
  (\bibinfo{year}{2007}).

\bibitem{kon07:11970}
\bibinfo{author}{\bibfnamefont{I.}~\bibnamefont{Kondov}},
  \bibinfo{author}{\bibfnamefont{M.}~\bibnamefont{Cizek}},
  \bibinfo{author}{\bibfnamefont{C.}~\bibnamefont{Benesch}},
  \bibinfo{author}{\bibfnamefont{H.}~\bibnamefont{Wang}}, \bibnamefont{and}
  \bibinfo{author}{\bibfnamefont{M.}~\bibnamefont{Thoss}}, \bibinfo{journal}{J.
  Phys. Chem. C} \textbf{\bibinfo{volume}{111}}(\bibinfo{number}{32}),
  \bibinfo{pages}{11970} (\bibinfo{year}{2007}).

\bibitem{tho07:153313}
\bibinfo{author}{\bibfnamefont{M.}~\bibnamefont{Thoss}},
  \bibinfo{author}{\bibfnamefont{I.}~\bibnamefont{Kondov}}, \bibnamefont{and}
  \bibinfo{author}{\bibfnamefont{H.}~\bibnamefont{Wang}},
  \bibinfo{journal}{Phys. Rev. B} \textbf{\bibinfo{volume}{76}},
  \bibinfo{pages}{153313} (\bibinfo{year}{2007}).

\bibitem{cra07:144503}
\bibinfo{author}{\bibfnamefont{I.~R.} \bibnamefont{Craig}},
  \bibinfo{author}{\bibfnamefont{M.}~\bibnamefont{Thoss}}, \bibnamefont{and}
  \bibinfo{author}{\bibfnamefont{H.}~\bibnamefont{Wang}}, \bibinfo{journal}{J.
  Chem. Phys.} \textbf{\bibinfo{volume}{127}}, \bibinfo{pages}{144503}
  (\bibinfo{year}{2007}).

\bibitem{wan08:139}
\bibinfo{author}{\bibfnamefont{H.}~\bibnamefont{Wang}} \bibnamefont{and}
  \bibinfo{author}{\bibfnamefont{M.}~\bibnamefont{Thoss}},
  \bibinfo{journal}{Chem. Phys.} \textbf{\bibinfo{volume}{347}},
  \bibinfo{pages}{139} (\bibinfo{year}{2008}).

\bibitem{wan08:115005}
\bibinfo{author}{\bibfnamefont{H.}~\bibnamefont{Wang}} \bibnamefont{and}
  \bibinfo{author}{\bibfnamefont{M.}~\bibnamefont{Thoss}},
  \bibinfo{journal}{New J. Phys.} \textbf{\bibinfo{volume}{10}},
  \bibinfo{pages}{115005} (\bibinfo{year}{2008}).

\bibitem{ego08:214303}
\bibinfo{author}{\bibfnamefont{D.}~\bibnamefont{Egorova}},
  \bibinfo{author}{\bibfnamefont{M.~F.} \bibnamefont{Gelin}},
  \bibinfo{author}{\bibfnamefont{M.}~\bibnamefont{Thoss}},
  \bibinfo{author}{\bibfnamefont{H.}~\bibnamefont{Wang}}, \bibnamefont{and}
  \bibinfo{author}{\bibfnamefont{W.}~\bibnamefont{Domcke}},
  \bibinfo{journal}{J. Chem. Phys.} \textbf{\bibinfo{volume}{129}},
  \bibinfo{pages}{214303} (\bibinfo{year}{2008}).

\bibitem{vel08:325}
\bibinfo{author}{\bibfnamefont{K.~A.} \bibnamefont{Velizhanin}},
  \bibinfo{author}{\bibfnamefont{H.}~\bibnamefont{Wang}}, \bibnamefont{and}
  \bibinfo{author}{\bibfnamefont{M.}~\bibnamefont{Thoss}},
  \bibinfo{journal}{Chem. Phys. Lett.} \textbf{\bibinfo{volume}{460}},
  \bibinfo{pages}{325} (\bibinfo{year}{2008}).

\bibitem{vel09:094109}
\bibinfo{author}{\bibfnamefont{K.~A.} \bibnamefont{Velizhanin}}
  \bibnamefont{and} \bibinfo{author}{\bibfnamefont{H.}~\bibnamefont{Wang}},
  \bibinfo{journal}{J. Chem. Phys.} \textbf{\bibinfo{volume}{131}},
  \bibinfo{pages}{094109} (\bibinfo{year}{2009}).

\bibitem{wan10:78}
\bibinfo{author}{\bibfnamefont{H.}~\bibnamefont{Wang}} \bibnamefont{and}
  \bibinfo{author}{\bibfnamefont{M.}~\bibnamefont{Thoss}},
  \bibinfo{journal}{Chem. Phys.} \textbf{\bibinfo{volume}{370}},
  \bibinfo{pages}{78} (\bibinfo{year}{2010}).

\bibitem{zho12:581}
\bibinfo{author}{\bibfnamefont{Y.}~\bibnamefont{Zhou}},
  \bibinfo{author}{\bibfnamefont{J.}~\bibnamefont{Shao}}, \bibnamefont{and}
  \bibinfo{author}{\bibfnamefont{H.}~\bibnamefont{Wang}},
  \bibinfo{journal}{Mol. Phys.} \textbf{\bibinfo{volume}{110}},
  \bibinfo{pages}{581} (\bibinfo{year}{2012}).

\bibitem{wan12:22A504}
\bibinfo{author}{\bibfnamefont{H.}~\bibnamefont{Wang}} \bibnamefont{and}
  \bibinfo{author}{\bibfnamefont{S.}~\bibnamefont{Shao}}, \bibinfo{journal}{J.
  Chem. Phys.} \textbf{\bibinfo{volume}{137}}, \bibinfo{pages}{22A504}
  (\bibinfo{year}{2012}).

\bibitem{man08:164116}
\bibinfo{author}{\bibfnamefont{U.}~\bibnamefont{Manthe}}, \bibinfo{journal}{J.
  Chem. Phys.} \textbf{\bibinfo{volume}{128}}, \bibinfo{pages}{164116}
  (\bibinfo{year}{2008}).

\bibitem{man09:054109}
\bibinfo{author}{\bibfnamefont{U.}~\bibnamefont{Manthe}}, \bibinfo{journal}{J.
  Chem. Phys.} \textbf{\bibinfo{volume}{130}}, \bibinfo{pages}{054109}
  (\bibinfo{year}{2009}).

\bibitem{Thoss04b}
\bibinfo{author}{\bibfnamefont{M.}~\bibnamefont{Thoss}},
  \bibinfo{author}{\bibfnamefont{I.}~\bibnamefont{Kondov}}, \bibnamefont{and}
  \bibinfo{author}{\bibfnamefont{H.}~\bibnamefont{Wang}},
  \bibinfo{journal}{Chem. Phys.} \textbf{\bibinfo{volume}{304}},
  \bibinfo{pages}{169} (\bibinfo{year}{2004}).

\bibitem{Li15}
\bibinfo{author}{\bibfnamefont{J.}~\bibnamefont{Li}},
  \bibinfo{author}{\bibfnamefont{H.}~\bibnamefont{Wang}}, \bibnamefont{and}
  \bibinfo{author}{\bibfnamefont{M.}~\bibnamefont{Thoss}},
  \bibinfo{journal}{J.\ Phys.: Condens.\ Matter} \textbf{\bibinfo{volume}{27}},
  \bibinfo{pages}{134202} (\bibinfo{year}{2015}).

\bibitem{Ness05}
\bibinfo{author}{\bibfnamefont{H.}~\bibnamefont{Ness}} \bibnamefont{and}
  \bibinfo{author}{\bibfnamefont{A.}~\bibnamefont{Fisher}},
  \bibinfo{journal}{Proc. Natl. Acad. Sci. USA} \textbf{\bibinfo{volume}{102}},
  \bibinfo{pages}{8826} (\bibinfo{year}{2005}).

\bibitem{Benesch06}
\bibinfo{author}{\bibfnamefont{C.}~\bibnamefont{Benesch}},
  \bibinfo{author}{\bibfnamefont{M.}~\bibnamefont{Cizek}},
  \bibinfo{author}{\bibfnamefont{M.}~\bibnamefont{Thoss}}, \bibnamefont{and}
  \bibinfo{author}{\bibfnamefont{W.}~\bibnamefont{Domcke}},
  \bibinfo{journal}{Chem. Phys. Lett.} \textbf{\bibinfo{volume}{430}},
  \bibinfo{pages}{355} (\bibinfo{year}{2006}).

\end{thebibliography}
\end{document}